\documentclass[A4paper,11pt]{article}
\usepackage{graphics,graphicx,epsfig,wrapfig}
\usepackage{longtable}  
\usepackage{amsmath,verbatim,amssymb,color,lscape}
\usepackage{natbib}
\usepackage{multirow} 
\usepackage{authblk}
\usepackage{bm}
\usepackage{indentfirst}
\usepackage[ruled,vlined]{algorithm2e}
\usepackage{algorithmic}
\usepackage[normalem]{ulem}
\usepackage[export]{adjustbox}
\usepackage{booktabs}
\usepackage{tabularx}
\usepackage{makecell}
\usepackage{hyperref}
\usepackage{verbatim}
\usepackage{url}
\usepackage{xr-hyper}
\newcommand{\hide}[1]{}

\makeatletter

\usepackage[outerbars,color]{changebar}
\ifx\pdfoutput\undefined
\else\ifnum\pdfoutput>0
  \usepackage{pdfcolmk}
\fi\fi
\cbcolor{black}

\usepackage[margin=1.0in]{geometry}
\usepackage{tikz}
\usetikzlibrary{bayesnet}
\usetikzlibrary{fit,positioning}

\newfont{\rmm}{cmr10 at 11pt}
\rmm

\title{Issue-Specific Polarization and Cohesion in a Multi-Party Legislature: Integrating the Latent Space Item Response Model with Topic-Based Regression}

\author[1]{Seungju Lee}
\author[1]{In-Kyun Kim}
\author[1,2]{Ick Hoon Jin}
\affil[1]{Department of Statistics and Data Science, Yonsei University. Seoul. Republic of Korea.}
\affil[2]{Department of Applied Statistics, Yonsei University. Seoul. Republic of Korea.}
\date{}

\begin{document}

\begin{titlepage}
\centering

\vspace*{2cm}

{\LARGE \textbf{Issue-Specific Polarization and Cohesion in a Multi-Party Legislature: Integrating the Latent Space Item Response Model with Topic-Based Regression}}

\vspace{2cm}

Seungju Lee$^1$, In-Kyun Kim$^1$, and Ick Hoon Jin$^{1,2,*}$

\vspace{1.5cm}

$^1$Department of Statistics and Data Science, Yonsei University, Seoul, Republic of Korea

$^2$Department of Applied Statistics, Yonsei University, Seoul, Republic of Korea

\vspace{2cm}

$^*$Corresponding Author:\\
Ick Hoon Jin\\
Department of Applied Statistics,\\ 
Department of Statistics and Data Science,\\
Yonsei University\\
50 Yonsei-ro, Seodaemun-gu, Seoul 03722, Republic of Korea\\
e-mail: ijin@yonsei.ac.kr

\vspace{2cm}

\textbf{Short Running Head:} Issue-Specific Polarization and Cohesion in a Multi-Party Legislature

\end{titlepage}

\maketitle

\begin{abstract}
We develop a \textcolor{black}{two-stage, cut-posterior} Bayesian framework for quantifying issue-specific legislative alignment in multi-party systems. The approach integrates a Latent Space Item Response Model (LSIRM), embedding legislators and bills in a shared Euclidean space, with Bayesian beta regression using text-derived topic proportions as bill-level covariates. The resulting legislator- and issue-specific coefficients allow polarization and cohesion to be compared across policy domains. \textcolor{black}{The beta-regression layer must not feed back into the estimated latent geometry, so that the two components target a cut posterior. We estimate it with a two-stage Multiple-Imputation procedure that propagates uncertainty in the latent positions into every downstream quantity.} In the 17th Korean National Assembly, fiscal domains such as Taxation and Grants and Local Government Budget show sharp polarization with tight within-party clustering, whereas Armed Services, Patriots, and Veterans exhibits weak party structuring and greater intra-party variability. The Democratic Labor Party forms a distinct cluster on several issues even where the two major parties are not strongly polarized, showing that legislative conflict escapes a single left--right ordering. The framework supports analysis of issue-structured voting in legislatures where one-dimensional ideal point models are unreliable.
\end{abstract}

\newpage

\section{Introduction}
\label{sec:introduction}

Ideal point estimation provides a statistical framework for representing high-dimensional roll-call voting data in a low-dimensional latent space, where each legislator is assigned a position that summarizes their revealed policy preferences. Conventional methods such as nominal three-step estimation \citep[NOMINATE;][]{Poole:1985, Poole:1987, Poole:1991, Poole:2007, Poole:1999, Poole:2005, wnominate2011, Carroll:2009, Carroll2009dw, Carroll:2013, voteview2025} and Bayesian item response theory \citep[BIRT;][]{Jackman:2001, pscl2024, Clinton2004, Clinton:2006, Clinton:2009} were developed for the United States Congress, where voting is primarily structured by a stable two-party system. Because most variation in U.S. roll-call behavior aligns with a single left--right ideological divide, one-dimensional ideal point estimates have become the standard representation in empirical research.

\begin{figure}[htb]
    \centering
    \begin{minipage}{0.47\linewidth}
        \centering
        \includegraphics[width=\linewidth]{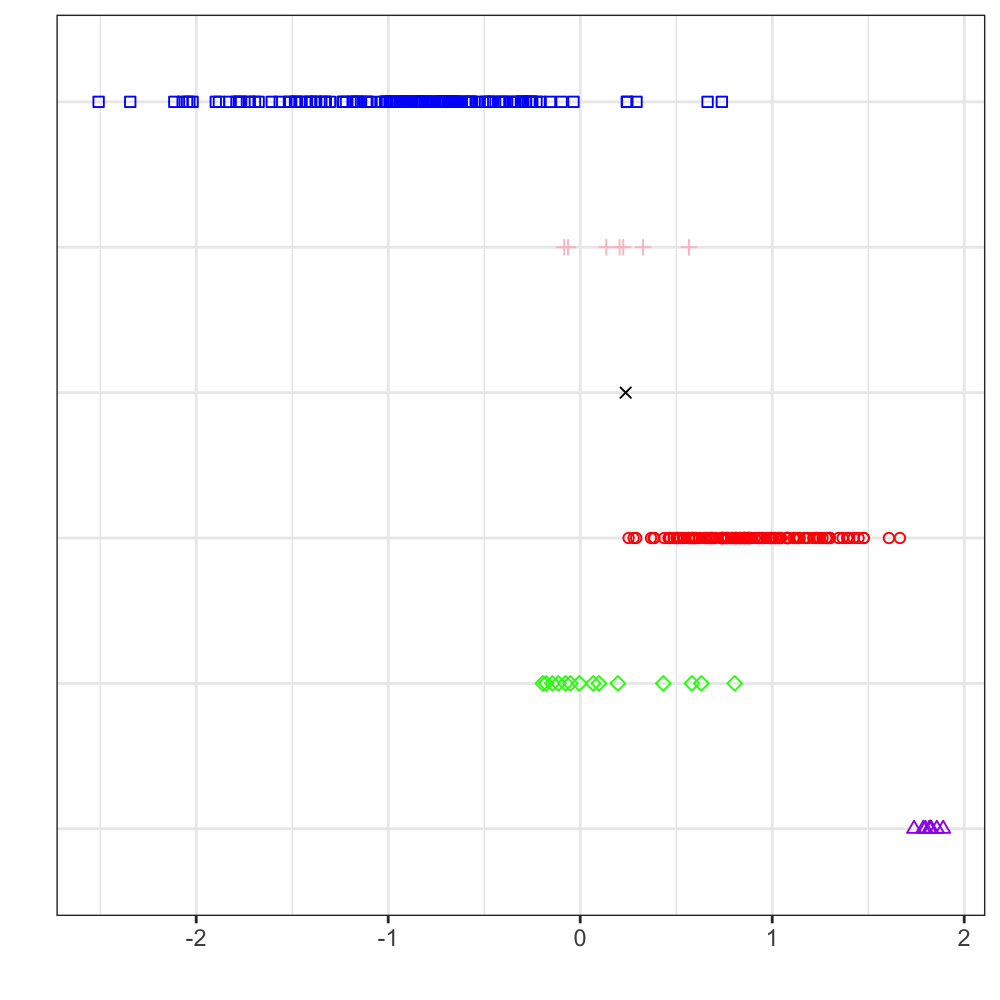}
        \text{(a) Bayesian IRT}
    \end{minipage}
    \begin{minipage}{0.47\linewidth}
        \centering
        \includegraphics[width=\linewidth]{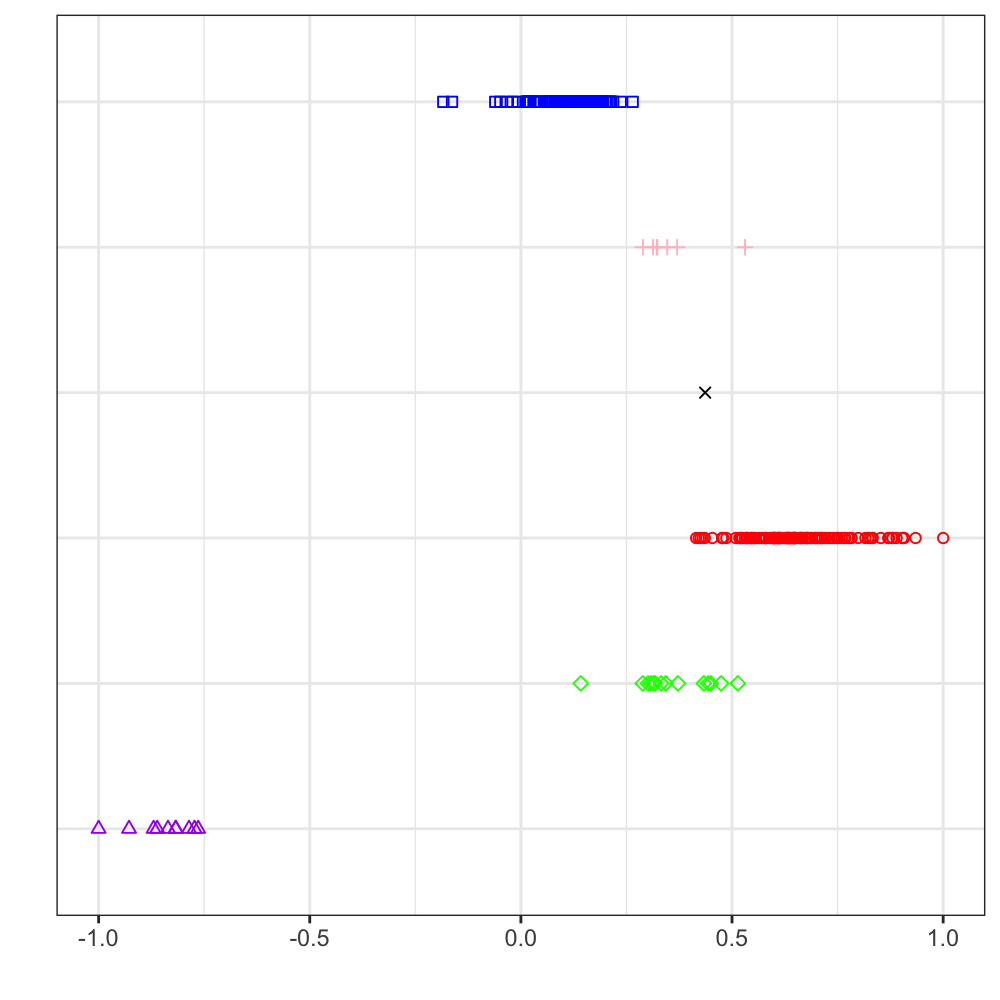}
        \text{(b) NOMINATE}  
    \end{minipage}
    \caption{One-dimensional ideal point estimates for legislators in the 17th Korean National Assembly under two conventional methods, Bayesian IRT (left) and NOMINATE (right). Each dot represents a legislator and is coded by party affiliation: Uri Party (blue squares), Grand National Party (red circles), Democratic Party (green diamonds), Democratic Labor Party (purple triangles), Liberal Democratic Union (pink plus signs), and Independents (black crosses).}
    \label{fig:17th_dim1_conventional}
\end{figure}

However, a one-dimensional summary that works well in the U.S. context does not generalize to many other legislatures \citep{hix2006dimensions}. The Korean National Assembly, for example, is a multi-party system in which ideological conflict varies substantially across issue areas \citep{rich2014party, kim2022democracy}. Even during periods dominated by two major blocs---such as the Uri Party (hereafter UP, liberal democrats) and the Grand National Party (hereafter GNP, conservatives) in the 17th National Assembly---smaller parties such as the Democratic Party (hereafter DP, conservative liberal democrats based in Jeolla province), the Liberal Democratic Union (hereafter LDU, conservatives based in Chungcheong province), and the Democratic Labor Party (hereafter DLP, social democrats) maintained a meaningful legislative presence and introduced distinct ideological cleavages \citep{shin2017legislative, jin2025anti}. Voting coalitions in Korea reconfigure across policy domains: issues such as economic growth, redistribution, labor rights, national security, and foreign affairs activate different cleavages, leading parties to align on some agendas while diverging on others. As a result, the structure of legislative conflict is inherently multidimensional, and one-dimensional ideal point models produce unstable or contradictory estimates when applied to this environment. \textcolor{black}{Figure~\ref{fig:17th_dim1_conventional}} illustrates this instability: under NOMINATE, UP members appear concentrated near the center while DLP legislators lie on the far left; under BIRT, UP members shift to the leftmost position and DLP legislators are pushed to the right. In contrast, GNP, LDU, and DP remain clustered on the right under both methods. The inconsistency in the estimated positions of UP and DLP demonstrates that conventional one-dimensional ideal point models do not yield a stable or interpretable ideological ordering in a multi-party legislature.

To address this challenge, we shift the analytical focus from interpreting latent dimensions to explaining why and how legislators diverge. The central idea is to integrate bill-level information directly into the modeling framework. Rather than attempting to interpret each latent axis as an ideological direction, we use observable characteristics of bills---such as committee labels or text-derived topic distributions---to understand how issue content shapes voting patterns. By explicitly incorporating the substantive content of bills, we obtain an issue-specific representation of legislative behavior that is more informative than a single global ideological scale.

Prior work on issue-specific ideal points has taken two main forms. The first approach relies on manually coded or externally assigned issue categories to partition roll-call data \citep{Moser:2021, shin2024measuring}, estimating separate ideal point models for each issue subset. The second approach integrates topic models with item response theory \citep{gerrish2010ideal, Gerrish:2012a, Gerrish:2012b, Lauderdale:2014}, using text-derived topics to adjust one-dimensional ideal points. These methods enrich the interpretation of ideal points but still treat legislator positions as the primary object of interest and do not directly model the geometry of legislator--bill interactions. In contrast, our method first estimates the continuous legislator--bill relationship in Euclidean space and then uses regression on bill-level issue features to explain how specific issues structure legislative conflict. This unifies categorical and continuous measures of issue content and provides a flexible, statistically principled framework for studying multi-party legislatures.

Our strategy has two main components. First, we model roll-call data as a bipartite network between legislators and bills and apply the latent space item response model \citep[LSIRM;][]{jeon:2020, lsirm2025, lee2025euclidean} to embed both sets of actors into a shared low-dimensional Euclidean space. LSIRM yields continuous latent distances $d_{ij}$ between legislator $i$ and bill $j$, providing a geometric representation of legislative conflict. \textcolor{black}{Representing this conflict in more than one latent dimension is not by itself unique to LSIRM: conventional ideal-point models such as NOMINATE and Bayesian IRT are also extensible to multiple dimensions \citep{luque2025operationalizing}. What distinguishes LSIRM here is not the number of latent dimensions but the interaction structure: legislators and bills are embedded jointly in the same space and related through a distance, rather than legislator positions being estimated from an inner-product interaction with bill-specific discrimination and difficulty parameters.} Second, we transform these distances into continuous affinity measures on $(0,1)$ and model them using beta regression, with bill-level covariates as predictors. The covariates include topic mixture proportions obtained from BERTopic \citep{Grootendorst:2022} applied to bill abstracts, together with additional categorical indicators for bill characteristics. The resulting regression coefficients are legislator- and issue-specific, summarizing how strongly each legislator tends to align with bills emphasizing a given issue area. \textcolor{black}{This paper builds on recent work that develops and validates Euclidean LSIRM for roll-call scaling \citep{lee2025euclidean} and extends it by using LSIRM-implied legislator--bill distances as inputs to a topic-based beta-regression layer, targeting a cut posterior estimated via a two-stage Multiple-Imputation procedure described in Section~\ref{sec:method}.}

This integrated design enables two complementary analyses. At the \emph{between-issue} level, we can identify which issue areas produce the greatest overall dispersion in legislator--bill affinities, highlighting topics that generate strong polarization or broad consensus. At the \emph{within-issue} level, we can measure inter-party division and intra-party heterogeneity for each issue by examining the distribution of legislator-specific coefficients across and within parties. Because LSIRM provides distances $d_{ij}$ from a single underlying geometric representation and all bills share a common issue-covariate vector $\mathbf{x}_j$, both cross-issue and within-issue comparisons are coherent within the same statistical framework.

\usetikzlibrary{positioning,calc}

\begin{figure}[htb]
\centering

\begin{tikzpicture}[
    node distance=1.2cm,
    every node/.style={font=\small},
    box/.style={rectangle, draw, rounded corners, align=center, minimum width=3.5cm, minimum height=1cm},
    smallbox/.style={rectangle, draw, rounded corners, align=center, minimum width=3.0cm, minimum height=0.8cm},
    arrow/.style={->, thick}
]

\node[box] (rollcall) {Roll-call Votes \\ (Legislator $\times$ Bill)};
\node[smallbox, below=of rollcall] (lsirm) {LSIRM \\ Euclidean Latent Space};
\node[smallbox, below=of lsirm] (dist) {Distances};

\node[box, right=5cm of rollcall] (text) {Bill Abstracts \\ (Korean Text)};
\node[smallbox, below=of text] (bertopic) {BERTopic \\ (Embeddings + Clustering)};
\node[smallbox, below=of bertopic] (topic) {Topic Proportions};

\node[box, below=2.2cm of $(dist)!0.5!(topic)$] (beta) {Beta Regression \\ $t_{ij} \sim \mathbf{X}_j$};
\node[smallbox, below=of beta] (coef) {Issue-Specific Coefficients \\ for each legislator};

\draw[arrow] (rollcall) -- (lsirm);
\draw[arrow] (lsirm) -- (dist);

\draw[arrow] (text) -- (bertopic);
\draw[arrow] (bertopic) -- (topic);

\draw[arrow] (dist) |- (beta);
\draw[arrow] (topic) |- (beta);

\draw[arrow] (beta) -- (coef);

\end{tikzpicture}
\caption{
Workflow of \textcolor{black}{the proposed LSIRM--beta-regression framework.} LSIRM maps roll-call votes to legislator--bill distances \(d_{ij}\), which are transformed into affinities \(t_{ij}\in(0,1)\). BERTopic extracts topic proportions \(\mathbf{X}_j\) from bill abstracts. The beta-regression component links \(t_{ij}\) to \(\mathbf{X}_j\), yielding legislator- and issue-specific coefficients \(\beta_{ik}\).
}
\end{figure}
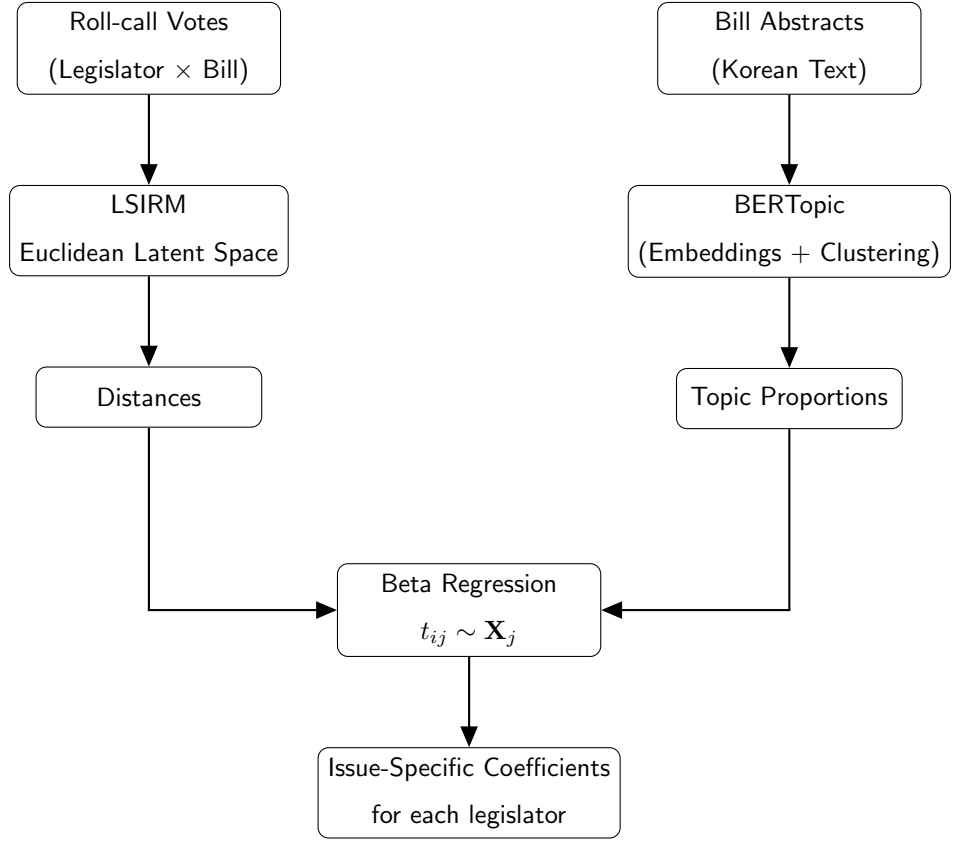

We apply this framework to roll-call votes and bill texts from the 17th National Assembly of the Republic of Korea (2004--2008), a \textcolor{black}{multi-party} legislature with a rich, multidimensional ideological structure. Our analysis reveals issue-dependent patterns of polarization and cohesion: some issue areas display clear separation between major parties, whereas others show substantial overlap and unexpected alignments involving smaller parties such as the DLP and LDU. More broadly, the results demonstrate that combining LSIRM with topic-based beta regression produces topic-specific ideological maps that are particularly informative in \textcolor{black}{multi-party} systems where conventional one-dimensional models yield unreliable estimates.

The remainder of this paper is organized as follows. Section~\ref{sec:method} introduces the methodological framework, Section~\ref{sec:data} describes the data and topic modeling procedure, Section~\ref{sec:results} presents the empirical findings for the 17th National Assembly, and Section~\ref{sec:conclusion} concludes by discussing implications for the study of issue-structured legislative behavior in \textcolor{black}{multi-party} systems and outlining avenues for future research. \textcolor{black}{Implementation details, extended diagnostic and validation checks, comparisons with alternative model specifications, and robustness analyses are reported in the Supplementary Material.}

\section{Method}\label{sec:method}
\subsection{Latent Space Item Response Model}\label{sec:lsirm}

Roll-call voting data can be viewed as a bipartite network between legislators and bills, where edges represent observed voting decisions. We use the latent space item response model \citep[LSIRM;][]{jeon:2020, lsirm2025} to embed legislators and bills in a shared Euclidean space and to obtain dyad-specific legislator--bill distances. Recent work establishes why enforcing a Euclidean (metric) geometry is consequential for distance-based representations of roll-call behavior \citep{lee2025euclidean}; here we take that geometric foundation as given and focus on integrating LSIRM-derived distances with bill-level issue information to quantify issue-specific alignment within a unified Bayesian framework.

Let $N$ denote the number of legislators and $P$ the number of bills. For legislator $i \in \{1,\dots,N\}$ and bill $j \in \{1,\dots,P\}$, let $y_{ij} \in \{0,1\}$ indicate whether legislator $i$ casts a yea vote on bill $j$ ($y_{ij} = 1$) or not ($y_{ij}=0$). LSIRM assumes a logistic response model of the form
\begin{equation}\label{lsirm}
\textcolor{black}{\mathrm{logit}}\big(P(y_{ij}=1 \mid a_i,b_j,\gamma,\mathbf{z}_i,\mathbf{w}_j)\big)
= a_i + b_j + g(\mathbf{z}_i,\mathbf{w}_j),
\end{equation}
where $a_i$ is legislator $i$'s general propensity to support bills (a legislator-specific intercept), $b_j$ is a bill-specific intercept capturing the baseline difficulty or popularity of bill $j$, and $g(\mathbf{z}_i,\mathbf{w}_j)$ is a distance-based effect linking latent positions to voting behavior. The vectors $\mathbf{z}_i \in \mathbb{R}^S$ and $\mathbf{w}_j \in \mathbb{R}^S$ denote the latent coordinates of legislator $i$ and bill $j$, respectively, in an $S$-dimensional Euclidean space.

We define $g(\mathbf{z}_i,\mathbf{w}_j)$ using the Euclidean distance
\[
d(\mathbf{z}_i,\mathbf{w}_j) = \lVert \mathbf{z}_i - \mathbf{w}_j \rVert_s,
\]
and specify
\[
g(\mathbf{z}_i,\mathbf{w}_j) = -\gamma\, d(\mathbf{z}_i,\mathbf{w}_j),
\]
where $\gamma > 0$ is a scale parameter governing how strongly distance in the latent space affects the voting probability. Under this specification, legislators are more likely to vote yea on bills that are close to them in the latent space, and the effect of distance is monotone and smooth. Since the likelihood depends only on pairwise distances, the latent coordinates are not identifiable up to rigid transformations; we address this by centering and aligning posterior draws in a post-processing step with Procrustes matching \citep{gower_generalized_1975}, as is standard in latent space models \citep{Hoff:2002, Shortreed:2006}.

We specify the following priors for LSIRM:
\begin{equation*}\label{eq:prior_lsirm}
\begin{split}
a_i &\sim N(0, \sigma_{a}^2), \quad \sigma_{a}^2 \sim \text{Inv-Gamma}(a_{\sigma}, b_{\sigma}), \quad \mathbf{z}_i \sim \text{MVN}_S(\mathbf{0}, \mathbf{I}_S),\\
b_j &\sim N(0, \sigma_{b}^2), \quad
\mathbf{w}_j \sim \text{MVN}_S(\mathbf{0}, \mathbf{I}_S), \quad \log \gamma \sim N(\mu_{\gamma}, \sigma_{\gamma}^2),
\end{split}
\end{equation*}
\textcolor{black}{where $\sigma_a^2$ is assigned a hierarchical Inv-Gamma$(a_{\sigma},b_{\sigma})$ prior and updated within the sampler, whereas $\sigma_b^2$ is fixed at a pre-specified value rather than estimated hierarchically; $(\mu_\gamma,\sigma_\gamma^2)$ are likewise fixed hyperparameters of the log-normal prior on $\gamma$. Numeric values for all LSIRM and beta-regression prior hyperparameters are reported in Supplementary Materials Section~\ref{sec:si_prior_specification}, together with the MCMC settings (iterations, burn-in, thinning) and convergence diagnostics for the baseline chain in Section~\ref{sec:si_convergence_check}.} The joint posterior density is then given by
\begin{equation}\label{eq:posterior_lsirm}
    \begin{split}
    \pi(\mathbf{a}, \mathbf{b}, \gamma, \mathbf{Z}, \mathbf{W} \mid \mathbf{Y})
    &\propto \prod_{i=1}^{N}\prod_{j=1}^{P} P(Y_{ij}=y_{ij} \mid a_i, b_j, \gamma, \mathbf{z}_i, \mathbf{w}_j)\\
    &\times \pi(\gamma) \prod_{i=1}^{N}\pi(a_i) \prod_{j=1}^{P}\pi(b_j) \prod_{i=1}^{N}\pi(\mathbf{z}_i) \prod_{j=1}^{P}\pi(\mathbf{w}_j).
    \end{split}
\end{equation}

\subsubsection{Choice of Latent-Space Dimension}\label{subsec:dimension_choice}

The choice of latent dimension in LSIRM involves a trade-off between parsimony and flexibility, and is particularly consequential in a joint framework where latent distances directly enter the subsequent beta-regression layer through the distance \(d_{ij}\). Different dimensional specifications can therefore induce systematically different affinity structures, potentially affecting the estimation of issue-specific coefficients.

To select an appropriate latent dimension for the LSIRM layer, we estimated models with dimensions ranging from one to six and compared their fit using three complementary information criteria: the Bayesian Information Criterion (BIC), the Deviance Information Criterion (DIC), and the Widely Applicable Information Criterion (WAIC). \textcolor{black}{These information criteria are computed using the LSIRM vote likelihood, because the purpose of this comparison is to assess how well each latent dimension captures the observed roll-call voting structure. The topic-based beta-regression likelihood is not included in this dimensionality comparison, since that layer is used subsequently to interpret the issue content of the estimated legislator–bill affinities.}

\begin{table}[!htbp]
\centering
\caption{\textcolor{black}{Dimensionality-selection information criteria for the LSIRM vote likelihood}}
\label{tab:dimension_model_criteria}
\begin{tabular}{rrrrrr}
\toprule
Dimension & BIC & DIC & DIC MCSE & WAIC & WAIC SE \\
\midrule
1 & \textcolor{black}{56,844.9} & \textcolor{black}{37,912.9} & \textcolor{black}{5.51} & \textcolor{black}{39,482.6} & \textcolor{black}{358.24} \\
2 & \textcolor{black}{62,896.2} & \textcolor{black}{34,504.2} & \textcolor{black}{7.03} & \textcolor{black}{36,680.2} & \textcolor{black}{343.37} \\
3 & \textcolor{black}{71,283.6} & \textcolor{black}{33,431.6} & \textcolor{black}{14.74} & \textcolor{black}{36,454.1} & \textcolor{black}{344.49} \\
4 & \textcolor{black}{79,888.9} & \textcolor{black}{32,576.9} & \textcolor{black}{15.72} & \textcolor{black}{36,203.4} & \textcolor{black}{344.04} \\
5 & \textcolor{black}{88,506.0} & \textcolor{black}{31,733.9} & \textcolor{black}{10.43} & \textcolor{black}{35,924.8} & \textcolor{black}{343.13} \\
6 & \textcolor{black}{97,144.4} & \textcolor{black}{30,912.3} & \textcolor{black}{18.52} & \textcolor{black}{35,731.4} & \textcolor{black}{342.55} \\
\bottomrule
\end{tabular}
\vskip .5cm
\begin{minipage}{0.95\linewidth}
\footnotesize
\textcolor{black}{Notes: Information criteria are computed using the LSIRM vote likelihood, since dimensionality selection concerns the fit of the latent vote-space representation to the observed roll-call matrix. The topic-based beta-regression likelihood is not included. BIC is a plug-in criterion and therefore has no posterior Monte Carlo standard error. DIC MCSE is computed using posterior-draw batch means. WAIC SE is computed from pointwise LSIRM WAIC contributions.}
\end{minipage}
\end{table}

\textcolor{black}{The criteria in Table~\ref{tab:dimension_model_criteria} reflect the trade-off between parsimony and fit. BIC increases monotonically as the dimension grows, which reflects its penalty for the number of latent-position parameters. DIC and WAIC decrease as dimensionality increases over the fitted range, which indicates improved LSIRM vote-likelihood fit in higher-dimensional latent spaces.}

\textcolor{black}{The largest improvement in both DIC and WAIC occurs when moving from one to two dimensions. WAIC decreases by 2802.4 from one to two dimensions, whereas adjacent-dimension improvements beyond two range from 193.4 to 278.6. DIC decreases by 3408.7 from one to two dimensions and improves by smaller amounts thereafter. The two-dimensional specification therefore captures the main improvement over the one-dimensional model while preserving direct visualization and interpretability of the latent voting space.}

\textcolor{black}{We use the two-dimensional LSIRM as the baseline specification for the main analysis. We interpret this choice as a parsimonious elbow specification rather than as the dimension that minimizes every information criterion. Higher-dimensional specifications continue to improve DIC and WAIC. Section~\ref{sec:si_dimension_beta_robustness} of the Supplementary Material shows that the issue-specific results obtained under three through six dimensions agree with the two-dimensional specification and with one another, whereas the one-dimensional specification does not. The consequential distinction is therefore between one dimension and two or more rather than among two through six dimensions, which supports selecting the most parsimonious specification within that band. Section~\ref{sec:si_dimension_simulation} of the Supplementary Materials reports a simulation study on how reliably this elbow criterion recovers the true latent dimension.}

\subsubsection{Advantages}

In the present framework, the value of LSIRM lies in the specific way it supports the issue-specific regression layer, rather than in re-establishing geometric properties of roll-call scaling. First, LSIRM yields a dyad-level continuous outcome---the legislator--bill distance $d_{ij}$---for every legislator--bill pair through a joint embedding of legislators and bills in a shared Euclidean space. These distances can be monotonically transformed into affinities and modeled as a function of bill-level covariates (e.g., topic proportions) that vary across bills but not across legislators. The broader implications of Euclidean metric structure for distance-based roll-call representations are discussed in \citet{lee2025euclidean}; here we leverage that structure to obtain well-behaved distances for downstream modeling.

Second, LSIRM includes legislator-specific intercepts $a_i$ that absorb heterogeneity in baseline approval tendencies (``yea bias''). This separation helps ensure that variation in $d_{ij}$ reflects content-related alignment rather than differences in legislators' overall propensity to support bills.

\subsection{Beta Regression for Issue-Specific Polarization and Cohesion}

\textcolor{black}{To study how issue content shapes legislative behavior we link the LSIRM geometry to bill-level covariates through beta regression \citep{ferrari2004beta}, the standard tool for modeling continuous responses on $(0,1)$, with a Bayesian implementation \citep{Jorge:2013}. LSIRM induces legislator--bill distances $d_{ij}$ which we map to affinities $t_{ij}\in(0,1)$. We model these affinities with a beta regression whose predictors $\mathbf{x}_j$ summarize bill issue content, including topic proportions from BERTopic and auxiliary bill indicators. The goal is to quantify for each legislator and each issue dimension how strongly that legislator aligns with bills emphasizing the dimension.}

\textcolor{black}{Given the LSIRM latent positions, we define the Euclidean distance $d_{ij} = \lVert \mathbf{z}_i - \mathbf{w}_j \rVert_2$ between legislator $i$ and bill $j$ and transform it into an affinity measure $t_{ij} = \exp(-d_{ij})$ with $0 < t_{ij} < 1$. Small distances give values of $t_{ij}$ close to $1$ and large distances give values near $0$. The support of $t_{ij}$ on $(0,1)$ matches the support of the beta distribution. The two beta parameters index a wide range of shapes on that support, from symmetric to left- or right-skewed and from diffuse to concentrated, which covers the affinity distributions observed across legislators. Section~\ref{sec:beta-distributional-diagnostics} of the Supplementary Material examines this fit.}

Following the Bayesian beta regression framework of \citet{Jorge:2013}, we assume that, for a given legislator $i$, the affinities $\{t_{ij}\}_{j=1}^P$ are conditionally independent draws from a beta distribution,
\[
t_{ij} \mid p_{ij}, q_{ij} \sim \textcolor{black}{\mathrm{Beta}}(p_{ij}, q_{ij}), \qquad 0 < t_{ij} < 1,
\]
with density
\[
\pi(t_{ij}; p_{ij},q_{ij}) = 
\frac{\Gamma(p_{ij}+q_{ij})}{\Gamma(p_{ij})\Gamma(q_{ij})}
t_{ij}^{\,p_{ij}-1}(1-t_{ij})^{\,q_{ij}-1},
\]
where $\Gamma(\cdot)$ denotes the gamma function. \textcolor{black}{We reparameterize the beta parameters in terms of the mean $\mu_{ij}$ and a legislator-specific precision $\phi_i$:
\[
\mu_{ij} = \frac{p_{ij}}{p_{ij}+q_{ij}}, \qquad \phi_i = p_{ij}+q_{ij},
\]
so that $p_{ij} = \mu_{ij}\phi_i$ and $q_{ij} = (1-\mu_{ij})\phi_i$. Under this parameterization,
\[
\mathbb{E}(t_{ij}) = \mu_{ij}, \qquad
\mathsf{Var}(t_{ij}) = \frac{\mu_{ij}(1-\mu_{ij})}{1+\phi_i},
\]
A larger $\phi_i$ indicates tighter concentration of the transformed affinities of legislator $i$ around the fitted mean. Because $\phi_i$ is legislator-specific rather than shared, this concentration differs across legislators.}

Let $\mathbf{x}_j$ denote the $K$-dimensional vector of bill-level covariates for bill $j$, including topic mixture proportions from BERTopic \citep{Grootendorst:2022} and additional indicators for other bill characteristics. Because the topic-proportion covariates sum to one within each bill, we use a reference-category parameterization in the regression layer: $\mathbf{x}_j$ includes an intercept and all but one topic-proportion covariate, so that coefficients are identified as deviations relative to an omitted baseline issue domain. For each legislator $i$, we model the conditional mean $\mu_{ij}$ via a logit link:

\begin{equation}\label{eq:beta_reg}
\textcolor{black}{\mathrm{logit}}(\mu_{ij}) = \mathbf{x}_j^{\top}\boldsymbol{\beta}_i,
\end{equation}
where $\boldsymbol{\beta}_i = (\beta_{i1},\dots,\beta_{iK})^{\top}$ is the vector of issue coefficients for legislator $i$. The corresponding likelihood is
\begin{equation}\label{eq:likelihood_cohesion}
    \begin{split}
        \mathcal{L}(\mathbf{t}|\boldsymbol{\beta}, \textcolor{black}{\boldsymbol\phi}) &= \prod_{i,j} P(t_{ij}|\boldsymbol{\beta}_i, \textcolor{black}{\phi_i})  \\
        &= \prod_{i,j} \frac{\Gamma(\textcolor{black}{\phi_i})}{\Gamma(\mu_{ij}\textcolor{black}{\phi_i})\Gamma((1-\mu_{ij})\textcolor{black}{\phi_i})} ({t}_{ij})^{\mu_{ij}\textcolor{black}{\phi_i}-1}(1-{t}_{ij})^{(1-\mu_{ij})\textcolor{black}{\phi_i}-1}
    \end{split}
\end{equation}
\textcolor{black}{The vector $\mathbf{x}_j$ omits one topic-proportion covariate as the reference category, as described in Section~\ref{sec:data}. A positive coefficient $\beta_{ik}$ therefore indicates that legislator $i$ shows higher affinity toward bills with larger values of covariate $k$ than toward bills dominated by the omitted reference topic. A negative coefficient indicates the opposite. The magnitudes and signs of $\boldsymbol{\beta}_i$ summarize the pattern of affinities of legislator $i$ across issues, each expressed relative to that same reference topic.}

\textcolor{black}{We allow the precision parameter $\phi_i$ of the beta-regression layer to vary by legislator rather than constraining it to a single value shared across all legislators. Comparability of the issue-specific coefficients $\boldsymbol{\beta}_i$ across legislators does not require a shared precision. The vector $\boldsymbol{\beta}_i$ enters the model only through the mean equation in Equation~\ref{eq:beta_reg}, which does not involve $\phi_i$. Every legislator is regressed on the same 471 bills, the same topic covariates $\mathbf{x}_j$, the same link function and the same reference category. These are the conditions under which coefficients from separately estimated regressions are treated as comparable, independent of how the residual-dispersion parameter is structured. Allowing $\phi_i$ to vary lets the model represent legislator-to-legislator heterogeneity in how tightly transformed affinities concentrate around the topic-driven mean, which a single shared precision cannot capture. Section~\ref{sec:variable-precision} of the Supplementary Material reports the evidence supporting this specification against a shared-precision alternative.}

\subsection{Estimation}

\textcolor{black}{We estimate the LSIRM parameters and the legislator-specific beta regressions within one Bayesian framework, with the two modules estimated in sequence. This subsection describes the joint posterior distribution and the two-stage cut-posterior Multiple-Imputation algorithm used to obtain posterior draws.}

\subsubsection{Joint Posterior}

Let $\Theta_{\mathrm{LS}} = \{\mathbf{a}, \mathbf{b}, \gamma, \mathbf{Z}, \mathbf{W}\}$ collect the LSIRM parameters (legislator intercepts, bill intercepts, scale parameter, and latent positions) and let $\Theta_{\mathrm{BR}} = \{\boldsymbol{B}, \textcolor{black}{\boldsymbol\phi}\}$ denote the issue-specific regression parameters, where $\boldsymbol{B}=\{\boldsymbol{\beta}_i\}_{i=1}^N$ contains legislator-level coefficient vectors and \textcolor{black}{$\boldsymbol\phi=\{\phi_i\}_{i=1}^N$ contains legislator-specific beta-regression precisions}. Here $\mathbf{Z}$ and $\mathbf{W}$ denote the $N \times S$ and $P \times S$ matrices whose rows are the latent-position vectors $\mathbf{z}_i$ and $\mathbf{w}_j$ defined in Section~\ref{sec:lsirm}.

\textcolor{black}{The latent positions $(\mathbf{Z},\mathbf{W})$ enter both modules of the model. They determine the roll-call likelihood $p(\mathbf{Y}\mid\Theta_{\mathrm{LS}})$ directly, and through the transformed distances $\mathbf{t}(\mathbf{Z},\mathbf{W})$ they also enter the beta-regression likelihood $p(\mathbf{t}(\mathbf{Z},\mathbf{W})\mid\boldsymbol{B},\boldsymbol\phi,\mathbf{X})$. We do not allow the second module to feed back into the estimation of $(\mathbf{Z},\mathbf{W})$. Because $\mathbf{t}(\mathbf{Z},\mathbf{W})$ is a deterministic function of the shared latent positions rather than an independently observed outcome, letting a possibly misspecified outcome family for $\mathbf{t}$ influence $(\mathbf{Z},\mathbf{W})$ would allow beta-regression fit quality to distort the latent ideological geometry that the roll-call votes alone are meant to identify. We therefore target a \emph{cut posterior} \citep{Plummer2015,Jacob2017}, which factorizes as
\begin{equation}
p_{\mathrm{cut}}(\Theta_{\mathrm{LS}},\Theta_{\mathrm{BR}} \mid \mathbf{Y},\mathbf{X}) \;:=\; p(\Theta_{\mathrm{LS}}\mid \mathbf{Y}) \; p(\Theta_{\mathrm{BR}} \mid \mathbf{t}(\mathbf{Z},\mathbf{W}),\mathbf{X}),
\label{eq:posterior_cut}
\end{equation}
where $\mathbf{t}(\mathbf{Z},\mathbf{W})$ denotes the transformed distances $t_{ij} = \exp\!\left(-\lVert \mathbf{z}_i - \mathbf{w}_j\rVert_2\right)$. The first factor
\[
p(\Theta_{\mathrm{LS}} \mid \mathbf{Y}) \;\propto\; p(\mathbf{Y}\mid\Theta_{\mathrm{LS}})\,p(\Theta_{\mathrm{LS}})
\]
is the posterior of the LSIRM module, computed from the roll-call likelihood alone, and the second factor
\[
p(\Theta_{\mathrm{BR}} \mid \mathbf{t}(\mathbf{Z},\mathbf{W}),\mathbf{X}) \;\propto\; p(\mathbf{t}(\mathbf{Z},\mathbf{W}) \mid \boldsymbol{B},\boldsymbol\phi,\mathbf{X})\,p(\Theta_{\mathrm{BR}})
\]
is the posterior of the beta-regression module conditional on the latent positions implied by $\Theta_{\mathrm{LS}}$, treating $\mathbf{t}(\mathbf{Z},\mathbf{W})$ as the data of that module. Equation~\eqref{eq:posterior_cut} is not the ordinary joint posterior of the two modules. It omits the contribution of the beta-regression likelihood to the conditional distribution of $\Theta_{\mathrm{LS}}$, which is what cutting the feedback path means.}

\textcolor{black}{The prior factorizes as $p(\Theta_{\mathrm{LS}})p(\Theta_{\mathrm{BR}})$, where $p(\Theta_{\mathrm{LS}})$ governs the latent-space parameters and is specified in Section~\ref{sec:lsirm}. For the regression layer, we use}
\begin{equation*}\label{eq:BR_prior}
    \begin{split}
        \boldsymbol{\beta}_i & \sim \mathcal{N}\big(\mathbf{0},\, \sigma_B^2 \mathbf{I}\big)\\
        \textcolor{black}{\phi_i} & \sim \mathrm{Gamma}(a_\phi, b_\phi)\textcolor{black}{, \quad i=1,\dots,N, \text{ independently}}
    \end{split}
\end{equation*}
\textcolor{black}{where $\boldsymbol{\beta}_i$ is assigned an independent, diffuse normal prior in the main specification, and each $\phi_i$ is assigned an independent Gamma prior with the same fixed hyperparameters $a_\phi,b_\phi$ across legislators -- the hyperparameters are shared, but the resulting $\phi_i$ are not pooled toward one another during estimation. As an alternative to this diffuse normal prior for $\boldsymbol{\beta}_i$, we separately evaluate a Zellner $g$-prior as a prior-sensitivity check rather than as the prior used to produce the main reported results (Supplementary Materials Section~\ref{sec:si_prior_gprior}). Numeric hyperparameter values for $\sigma_B$, $a_\phi$, and $b_\phi$ are reported in Supplementary Materials Section~\ref{sec:si_prior_specification}, together with the MCMC settings and convergence diagnostics for the baseline chain in Section~\ref{sec:si_convergence_check}.}

The likelihood factorizes into two components. For the roll-call votes:
\begin{equation*}\label{eq:likeli_LSIRM}
    p(Y_{ij}=1 \mid \Theta_{\mathrm{LS}}) =
    \textcolor{black}{\mathrm{logit}}^{-1}(a_i + b_j - \gamma \lVert \mathbf{z}_i - \mathbf{w}_j\rVert_2),
\end{equation*}
and for issue-specific polarization and cohesion, conditional on $(\mathbf{Z},\mathbf{W})$:
\begin{equation*}\label{eq:likeli_BR}
\begin{split}
    t_{ij} \mid \boldsymbol{\beta}_i,\textcolor{black}{\phi_i}
    & \sim \mathrm{Beta}(\mu_{ij}\textcolor{black}{\phi_i},(1-\mu_{ij})\textcolor{black}{\phi_i}), \\
    \qquad
    \mu_{ij} &=\textcolor{black}{\mathrm{logit}}^{-1}(\mathbf{x}_j^\top\boldsymbol{\beta}_i)
\end{split}
\end{equation*}

\subsubsection{\textcolor{black}{Cut-Posterior Estimation via Multiple Imputation}}

\textcolor{black}{Equation~\eqref{eq:posterior_cut} factorizes into two modules, and we estimate it in two matching stages. A first stage draws $\Theta_{\mathrm{LS}}$ from $p(\Theta_{\mathrm{LS}}\mid\mathbf{Y})$. A second stage draws $\Theta_{\mathrm{BR}}$ from $p(\Theta_{\mathrm{BR}}\mid\mathbf{t}(\mathbf{Z},\mathbf{W}),\mathbf{X})$ conditional on each of $m$ first-stage draws held fixed in turn, and pools the $m$ resulting posteriors. The two stages form a Multiple Imputation scheme \citep{Rubin1987}, in which the first-stage draws play the role of $m$ completed imputations of the latent positions. Pooling across them performs by explicit Monte Carlo integration the marginalization}
\begin{equation}\label{eq:mi_target}
p(\Theta_{\mathrm{BR}}\mid\mathbf{Y},\mathbf{X}) = \int p(\Theta_{\mathrm{BR}}\mid \mathbf{t}(\mathbf{Z},\mathbf{W}),\mathbf{X})\, p(\Theta_{\mathrm{LS}}\mid\mathbf{Y})\, d\Theta_{\mathrm{LS}}
\end{equation}
\textcolor{black}{that a single joint chain updating $\Theta_{\mathrm{BR}}$ conditional on its own evolving trajectory of $(\mathbf{Z},\mathbf{W})$ is not in general guaranteed to target \citep{Plummer2015}.}

\textcolor{black}{\emph{Stage 1: drawing $\Theta_{\mathrm{LS}}$.} We run a Metropolis--Hastings-within-Gibbs chain over $(\boldsymbol{a},\boldsymbol{b},\gamma,\mathbf{Z},\mathbf{W})$ using the roll-call likelihood alone, so that every retained draw is by construction a draw from $p(\Theta_{\mathrm{LS}}\mid\mathbf{Y})$. At iteration $l$:}
\begin{enumerate}
    \item \textcolor{black}{\textbf{Impute missing roll-call votes:} Each missing $Y_{ij}$ is sampled from the Bernoulli likelihood using the current latent positions.}
    \item \textcolor{black}{\textbf{Update LSIRM parameters} $(\boldsymbol{a},\boldsymbol{b},\gamma,\mathbf{Z},\mathbf{W})$: each component is updated by a random-walk Metropolis step using the full conditional likelihood of the roll-call data.}
    \item \textcolor{black}{\textbf{Update the hyperparameter} $\sigma_a^2$: it enters conjugately and is updated via its inverse-gamma conditional.}
\end{enumerate}
\textcolor{black}{After burn-in, we thin the resulting chain to obtain $m$ approximately independent draws of $\Theta_{\mathrm{LS}}$, selected using the autocorrelation of the implied legislator--bill distances $d_{ij}$ rather than the raw latent coordinates, which are inflated by the rotational non-identifiability of the model. This same chain also updates $\Theta_{\mathrm{BR}}=(\boldsymbol B,\boldsymbol\phi)$ at every iteration, conditional on the $(\mathbf{Z},\mathbf{W})$ of that iteration. We retain those values at each of the $m$ selected iterations as a warm start for Stage 2 rather than as part of the reported $\Theta_{\mathrm{BR}}$ posterior.}

\textcolor{black}{\emph{Stage 2: drawing $\Theta_{\mathrm{BR}}$ conditional on each fixed $\Theta_{\mathrm{LS}}$ draw.} For each of the $m$ draws, we hold $(\mathbf{Z},\mathbf{W})$ fixed and run an independent Metropolis--Hastings chain to convergence over $\Theta_{\mathrm{BR}}=(\boldsymbol B,\boldsymbol\phi)$ alone:}
\begin{enumerate}
    \item \textcolor{black}{\textbf{Compute transformed distances:} $t_{ij} = \exp\!\left(-\lVert \mathbf{z}_i - \mathbf{w}_j\rVert_2 \right)$ from the fixed latent positions. These are the data for the beta regressions below.}
    \item \textcolor{black}{\textbf{Update legislator-specific issue coefficients} $\boldsymbol{\beta}_i$: For each legislator a Metropolis step proposes a new coefficient vector, and the acceptance ratio uses the beta likelihood based on $t_{ij}$ together with the chosen prior.}
    \item \textcolor{black}{\textbf{Update legislator-specific precision parameters} $\phi_i$: For each legislator, $\phi_i$ is updated through a log-normal random-walk Metropolis step based on the beta likelihood of the transformed affinities $\{t_{ij}\}_{j=1}^P$ of that legislator.}
\end{enumerate}

\textcolor{black}{
Because each of the $m$ Stage 2 chains already yields draws from its own conditional posterior $p(\Theta_{\mathrm{BR}} \mid \Theta_{\mathrm{LS}}^{(l)}, \mathbf{t}, \mathbf{X})$, we do not reduce these draws to a mean and variance before combining them. We pool the retained post-burn-in draws from all $m$ chains into a single combined sample. Because the $m$ values of $\Theta_{\mathrm{LS}}^{(l)}$ are themselves approximately independent draws from $p(\Theta_{\mathrm{LS}}\mid\mathbf{Y})$, this combined sample is the Monte Carlo estimate of the integral in Equation~\eqref{eq:mi_target}, obtained by replacing it with a finite average over the $m$ draws. Sampling a chain index $l$ uniformly and then a draw from that chain is equivalent to sampling from the corresponding equally-weighted mixture. We report the mean and the 2.5th and 97.5th empirical percentiles of this pooled sample as the posterior mean and $95\%$ credible interval for every $\Theta_{\mathrm{BR}}$-derived quantity in the paper. Point estimates and credible intervals are ordinary posterior summaries of this pooled sample, with no distributional approximation involved.}

\textcolor{black}{For interpretability, we additionally report, for each scalar component $\theta$ of $\Theta_{\mathrm{BR}}$, the classical decomposition of the pooled variance into a within-draw component $\bar U=\frac{1}{m}\sum_l \hat U^{(l)}$ \citep{Rubin1987}, the average posterior variance of $\theta$ conditional on a single fixed $\Theta_{\mathrm{LS}}$ draw, and a between-draw component $B=\frac{1}{m-1}\sum_l(\hat\theta^{(l)}-\bar\theta)^2$, which measures how much the posterior mean $\hat\theta^{(l)}$ varies across the $m$ draws of $\Theta_{\mathrm{LS}}$, where $\hat\theta^{(l)}$ and $\hat U^{(l)}$ are the posterior mean and variance of $\theta$ within the $l$-th chain alone. Section~\ref{sec:si_mi_validation} of the Supplementary Material reports this decomposition. It quantifies how much of the total uncertainty in $\theta$ is attributable to not knowing $\Theta_{\mathrm{LS}}$ exactly, and it is a diagnostic rather than an input to the reported point estimates or intervals.}

\subsubsection{Posterior Output and Interpretation}

\textcolor{black}{We report $\Theta_{\mathrm{LS}}$-derived quantities, namely the latent legislator and bill positions and the LSIRM roll-call parameters, directly from the Stage 1 posterior. All issue-specific quantities derived from $\Theta_{\mathrm{BR}}=(\boldsymbol B,\boldsymbol\phi)$, namely legislator-level topic coefficients, party contrasts and within-party heterogeneity, are reported from the Stage 2 Multiple-Imputation-pooled posterior described above, which is our estimate of $p(\Theta_{\mathrm{BR}}\mid\mathbf{Y},\mathbf{X})$ in Equation~\eqref{eq:posterior_cut}. All posterior summaries, including credible intervals, density estimates and derived polarization measures, are computed from these samples. Section~\ref{sec:si_convergence_check} of the Supplementary Material reports additional MCMC convergence diagnostics, namely trace plots for the global scale parameter $\gamma$, representative legislator-specific precision parameters $\phi_i$ and the log posterior, together with representative latent positions and regression coefficients and the full Multiple-Imputation protocol. Section~\ref{sec:si_mi_validation} of the Supplementary Material reports the pooling diagnostics separately.}

\section{The 17th Korean National Assembly Data}\label{sec:data}

\subsection{The Dataset of Roll-call Voting}

We analyze roll-call data from the 17th National Assembly of the Republic of Korea (2004--2008), constructed by processing official PDF minutes provided by the National Assembly Secretariat \citep{NAK2024_record}. South Korea provides a compelling empirical setting because its legislature operates under a multi-party system shaped by multiple ideological cleavages. To quantify the degree of multipartism, we compute the effective number of parties, which accounts for both the number of parties and their relative seat shares. Using the \citet{laakso1979effective} index, we estimate that the effective number of seat parties during this term is 2.36. This value suggests that the 17th National Assembly was best characterized as a moderate multi-party---or ``two-and-a-half--party''---system, with a limited yet meaningful level of party competition.


The resulting data form an $N \times P$ binary matrix, where $N$ is the number of legislators and $P$ is the number of bills that reached a plenary vote. Each cell $y_{ij}$ records legislator $i$’s vote on bill $j$, and the original minutes classify votes into four categories: ``Yea,'' ``Nay,'' ``Abstention,'' and ``Absence.'' We code ``Yea'' as $1$ and both ``Nay'' and ``Abstention'' as $0$, so that $y_{ij} = 1$ indicates support for bill $j$ and $y_{ij} = 0$ indicates explicit opposition or non-support\textcolor{black}{; recorded ``Absence'' is treated as missing rather than as either support or opposition (see below).}

\textcolor{black}{The ``Absence'' category covers votes with no individual record. These arise from institutional nonparticipation rather than from a voting choice, and include holding certain parliamentary offices, membership in the executive branch and procedural ineligibility to vote on a particular bill. We set $y_{ij}$ to missing in these cases and treat it as an unobserved outcome in the LSIRM likelihood rather than as support or opposition.} \textcolor{black}{To assess the sensitivity of our results to this coding decision, we additionally conduct a robustness analysis that treats recorded abstentions as missing while retaining the same bill sample (Supplementary Materials Section~\ref{sec:si_abstention_coding}). We additionally evaluated an extension that models ``Yea,'' ``Nay,'' and ``Abstention'' as three distinct outcome categories rather than collapsing the latter two (Supplementary Materials Section~\ref{sec:si_multicategory_abstention}).}

\textcolor{black}{Bills in the Korean legislature require only a simple majority of members present to pass. The distribution of votes is nonetheless skewed. Across all bills that reached the floor in the 17th National Assembly the mean proportion of ``Yea'' votes is $96.1\%$. This pattern reflects institutional practice. Most substantive bargaining over the content of legislation occurs at the committee stage, and contested provisions are amended or withdrawn before bills are scheduled for a plenary vote. Bills that fail to reach the floor by the end of the legislative term are automatically nullified, which further reduces the number of divisive roll-calls.}

For our analysis, we focus on roll-calls that contain meaningful information about ideological disagreement. Following the existing literature \citep{Poole:1985, Poole:2007, voteview2025, Clinton2004, pscl2024}, bills with nearly unanimous approval or rejection---those with ``Yea'' rates below 0.025 or above 0.975---are excluded because they provide little information about ideological differentiation. Although such bills may be substantively important, they are not useful for identifying fine-grained patterns of legislative conflict in a spatial model. The resulting dataset therefore emphasizes substantive roll-calls in which legislators' ideological differences are empirically identifiable. \textcolor{black}{As a robustness check on this filtering choice, we additionally estimate the LSIRM vote layer using the full, unfiltered roll-call record (including near-unanimous bills) and confirm that the issue-specific results are not driven by this exclusion criterion (Supplementary Materials Section~\ref{sec:si_fullvote_indexed}).}

\textcolor{black}{Of the 1,936 bills appearing in the original legislative records, 494 remain after the roll-call availability and vote-variation criteria are applied. BERTopic assigns 23 of these bills to its outlier cluster with topic ID $-1$, and substantive topic-proportion covariates are not used for them in the regression layer. The joint LSIRM--beta regression model is estimated on the common analytic sample of 471 bills and 322 legislators. Table~\ref{tab:data_summary} summarizes the dimensions of this sample and Table~\ref{tab:topic_proportions} reports the distribution of these bills across topics.}

\begin{table}[htbp]
\centering
\begin{tabular}{lr}
\toprule
\textbf{Dimension} & \textbf{Value} \\
\midrule
Legislators ($N$) & 322 \\
\quad Uri Party (UP) & 155 \\
\quad Grand National Party (GNP) & 135 \\
\quad Democratic Party (DP) & 14 \\
\quad Democratic Labor Party (DLP) & 10 \\
\quad Liberal Democratic Union (LDU) & 7 \\
\quad Independent & 1 \\
Bills ($P$) & 471 \\
Substantive topics ($K$) & 8 \\
Missing legislator--bill entries & 34.91\% \\
Observed ``Yea'' proportion & 0.883 \\
\bottomrule
\end{tabular}
\caption{
\label{tab:data_summary}
Dimensions of the final analytic sample used in the joint LSIRM--beta regression model, including the distribution of legislators across parliamentary groups. Table~\ref{tab:topic_proportions} reports the distribution of these bills across topics.
}
\end{table}

\subsection{The Dataset of Bills}

In our framework, the interaction between legislators and bills is characterized not only by voting outcomes but also by the substantive content of each bill. To capture this content, we construct a bill-information matrix $\mathbf{X}$ whose rows index bills and whose columns represent issue-related features. Formally, $\mathbf{X}$ has dimension $P \times K$, where $P$ is the number of bills and $K$ is the number of issue covariates. Different constructions of $\mathbf{X}$ correspond to different ways of summarizing what each bill is about. For example, the columns of $\mathbf{X}$ can be manually coded policy indicators, committee-based dummies, or text-derived topic proportions. In this study, we focus on semantically defined topics obtained from bill abstracts and use committee-based indicators as auxiliary information.

Bill abstracts and metadata were collected from the legislative information system of the National Assembly \citep{NAK2024_bills}. Abstracts are concise summaries that describe the purpose, main provisions, and policy background of each bill, and they are written in a relatively standardized format. We use these abstracts rather than full bill texts because they capture the core policy content in a compact form and provide a consistent basis for text analysis across a large number of proposals.

\subsubsection{Topic Modeling}\label{subsec:topic_modeling}

The construction of the bill-information matrix $\mathbf{X}$ proceeds in three steps. As a first step, we determine the number of latent topics that will serve as issue dimensions. We apply the BERTopic framework to the corpus of bill abstracts and evaluate topic quality over a candidate range of topic counts $k \in \{2,\ldots,17\}$. The upper bound reflects the institutional structure of the 17th Korean National Assembly, which comprises 17 standing committees corresponding to major policy domains.

To assess robustness to stochastic initialization, we estimate BERTopic models under 500 independent random seeds for each candidate topic count $k\in\{2,\ldots,17\}$. Because BERTopic includes non-deterministic components, we fix the random state at each stage so that each $(k,\text{seed})$ pair yields a deterministic solution, resulting in 8{,}000 fitted models in total. For each fitted model, we compute two semantically motivated coherence metrics, $C_v$ and $C_{\text{NPMI}}$ \citep{roder2015exploring, newman2010automatic}. For a given metric and seed, we record the topic count $k$ that attains the highest coherence score and then summarize the empirical distribution of selected topic counts across seeds.

\begin{table}[htbp]
\centering
\small
\renewcommand{\arraystretch}{1.2}
\begin{tabular}{lcc}
\toprule
\textbf{Coherence Metric} & \textbf{Modal \(k\)} & \textbf{Interpretation} \\
\midrule
\(C_v\)              & 9 & Semantic coherence \\
\(C_{\text{NPMI}}\)  & 9 & Normalized semantic coherence \\
\bottomrule
\end{tabular}
\caption{
Modal topic count selected by each coherence metric across 500 random seeds.
Both $C_v$ and $C_{\text{NPMI}}$ exhibit a clear mode at $k=9$.
}
\label{tab:topic_coherence_summary}
\end{table}

Both coherence measures consistently select an interior topic granularity, with a clear mode at $k=9$ (Table~\ref{tab:topic_coherence_summary}). We therefore adopt $k=9$ as the primary specification for the main analysis, prioritizing $C_{\text{NPMI}}$ because it balances normalized co-occurrence structure with interpretability. All subsequent analyses use the $k=9$ solution corresponding to the highest $C_{\text{NPMI}}$ score among the 500 seeds. Additional coherence diagnostics and the full seed-by-$k$ selection distributions are reported in the Supplementary Materials (Section A.1; Figure A1).

With the number of topics fixed, the second step embeds bill texts into a semantic vector space using a Korean sentence-level language model (KR-SBERT; \citealp{kr-sbert}) as the underlying embedding model within BERTopic. Each abstract is mapped to a dense vector that summarizes its semantic content in context. BERTopic then applies dimensionality reduction (UMAP) followed by density-based clustering (HDBSCAN) to identify coherent clusters of semantically related texts. Unlike frequency-based topic models such as Latent Dirichlet Allocation (LDA), this embedding-based approach leverages semantic similarity in Korean legislative language, allowing bills that address similar policy areas to be grouped together even when they use different terminology.

Third, we refine the tokenization and vocabulary using Kiwi \citep{Kiwi}, a morphological analyzer for Korean. Kiwi performs part-of-speech tagging, lemmatization, and compound word segmentation tailored to Korean morphology, which reduces noise from inflected forms and improves the stability and interpretability of the resulting topics. We apply Kiwi-based preprocessing before fitting the final BERTopic model so that high-frequency function words and purely procedural terms do not dominate the learned topics.

\begin{table}[htbp]
\centering
\begin{tabular}{c|l}
\toprule
\textbf{Topics} & \textbf{Most related words} \\ 
\midrule
Topic 1 & bank, broadcast, industry, company, technology, research, business, hire, educate, invest \\ 
Topic 2 & planning, housing, development, business, urban, region, construction, facility, area, sites \\
Topic 3 & tax, deduction, income, house, transfer, levy, income tax, payment, report, tax amount \\
Topic 4 & coercion, mobilize, imperial, bereave, victim, investigate, occupy, examine, bury, nation \\
Topic 5 & lawyer, judge, jury, disciplinary, court, trial, judiciary, case, punishment, judiciary \\
Topic 6 & province, autonomy, resident, execute, local, ordinance, police, governor, claim, authority \\
Topic 7 & grant, local, subsidy, budget, decentralize, educate, resource, autonomy, concede, account \\
Topic 8 & bankrupt, sentence, restore, commit, right, medicate, discharge, medical, qualify, interest \\

\bottomrule
\end{tabular}
\caption{
\label{tab:topic-word}
Results of BERTopic applied to the abstracts of 494 voted bills. For each topic, we report the most representative terms that characterize its semantic content. These topics form the columns of the bill-information matrix $\mathbf{X}$ and provide a quantitative summary of how strongly each bill is associated with each issue domain.
}
\end{table}

\textcolor{black}{Table~\ref{tab:topic-word} summarizes the eight substantive topics identified from the abstracts of the 494 roll-call bills retained after the voting-data filters and lists the most representative words for each topic. In the BERTopic model-selection stage $k=9$ denotes eight substantive topic clusters together with one outlier cluster. The outlier cluster carries topic ID $-1$, contains 23 bills and is excluded from substantive interpretation and from the downstream topic-based regression. The remaining 471 bills are assigned to one of the eight substantive topics and constitute the common analytic sample used in the joint LSIRM--beta regression model.}

Based on close reading of representative bills and their key terms, we interpret these topics as Business (Topic 1), Land and Regional Development (Topic 2), Taxation (Topic 3), Armed Services, Patriots, and Veterans (Topic 4), Judiciary (Topic 5), Local Governmental Affairs (Topic 6), Grants and Local Government Budget (Topic 7), and Social (Topic 8). Because topic proportions sum to one within each bill, including all eight topic proportions together with an intercept would induce perfect collinearity. We therefore adopt a reference-category parameterization by omitting one topic proportion and including an intercept, so that coefficients represent deviations from the omitted baseline issue domain. In the main analysis, we omit the Business topic as the baseline, so the bill-information vector $\mathbf{x}_j$ consists of an intercept and seven topic-proportion covariates. \textcolor{black}{Because the topic proportions are compositional, identifying the regression requires omitting exactly one topic as a reference category. This is the same identification problem as the dummy-variable trap for categorical covariates. We select Business as the reference category because it has the largest average share among the eight topics in our sample, with a mean proportion of $0.291$ reported in Table~\ref{tab:topic_proportions}. Using the most prevalent topic as the omitted category is a common convention for compositional and categorical covariates. It tends to minimize the standard errors of the resulting contrasts and yields an interpretable baseline, since each coefficient is then expressed relative to the most typical bill composition in the sample. The corresponding coefficients therefore represent issue-affinity contrasts relative to Business. The choice of reference is a reparameterization. Fitted affinities and all between-topic contrasts are invariant to it, and only the party contrast reported for each topic shifts, by exactly the party contrast of the omitted topic. Business is also one of the least partisan topics in our sample by the same underlying vote record. Section~\ref{sec:si_reference_category_robustness} of the Supplementary Materials reports a $4.3\%$ UP--GNP majority disagreement rate on Business bills against $10.3\%$ across all 471 bills, so that the induced shift is small relative to the contrasts we report.}

\begin{table}[htb!]
\centering
\caption{\textcolor{black}{Distribution of continuous topic proportions across the 471 analytic-sample bills.}}
\label{tab:topic_proportions}
\begin{tabular}{lc}
\toprule
Topic & Mean proportion \\
\midrule
Business & 0.291 \\
Land and Regional Development & 0.200 \\
Taxation & 0.129 \\
Armed Services, Patriots, and Veterans & 0.040 \\
Judiciary & 0.046 \\
Local Governmental Affairs & 0.128 \\
Grants and Local Government Budget & 0.125 \\
Social & 0.041 \\
\midrule
Total & 1.000 \\
\bottomrule
\end{tabular}
\vskip .50cm
\begin{minipage}{0.92\linewidth}
\footnotesize
\textcolor{black}{\textit{Note:} Proportions sum to one within each bill, so that the column mean sums to one across topics. One topic typically dominates a given bill, and the largest topic accounts for a mean of $70.6\%$ of the estimated composition of a bill, with a median of $70.4\%$. Multiple topics are nonetheless co-active. Under an activation threshold of $0.05$, bills have a mean of $2.55$ active topics, a median of $2$ and a maximum of $6$. We therefore retain the full continuous proportions rather than a single dominant-topic label. Section~\ref{sec:si_topic_representation_robustness} of the Supplementary Materials reports this distribution under two discrete representations used for a robustness comparison.}
\end{minipage}
\end{table}

These topics, together with their corresponding proportions for each bill, constitute the issue covariates in our beta regression and allow us to link substantive issue content to \textcolor{black}{legislator-bill affinity} in the latent Euclidean space. The topic labels are used only to aid interpretation; all formal analyses use the continuous topic-proportion covariates rather than discrete topic assignments. Additional implementation details for Korean preprocessing, BERTopic settings, and the full seed-by-$k$ coherence selection distributions (including $C_{\text{UCI}}$ and $u$-mass) are provided in the Supplementary Materials (Sections A.1; Figure A1).

\section{Results}\label{sec:results}
\subsection{Multidimensional Structure of Legislative Behavior in the 17th Korean National Assembly}\label{subsec:multidim_structure}

\begin{figure}[htb]
    \centering
    \begin{minipage}{0.47\linewidth}
        \centering
        \includegraphics[width=\linewidth]{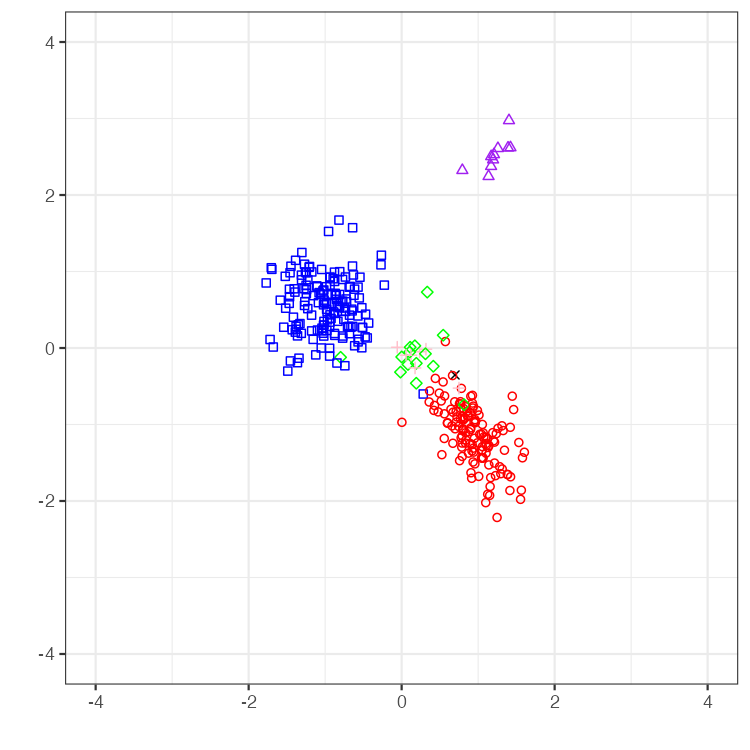}
        \text{Legislator map}
    \end{minipage}
    \begin{minipage}{0.47\linewidth}
        \centering
        \includegraphics[width=\linewidth]{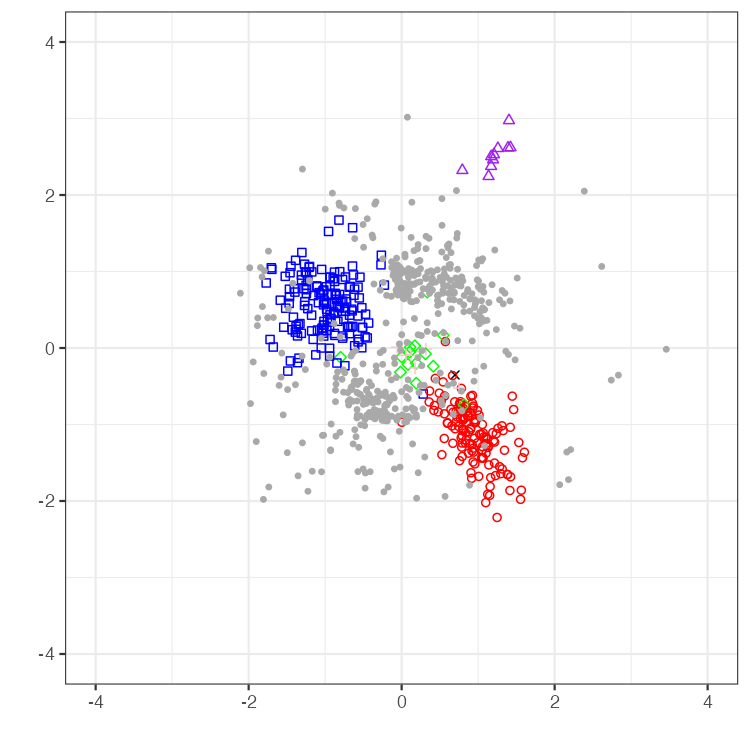}
        \text{Joint legislator--bill map}
    \end{minipage}
    \caption{
    Two-dimensional LSIRM configuration for the 17th Korean National Assembly. The left panel displays legislators only; the right panel overlays bills as gray points. Legislators are coded by party: UP (liberal democrats, blue squares), GNP (conservatives, red circles), DP (conservative liberal democrats based in Jeolla, green diamonds), DLP (social democrats, purple triangles), LDU (conservatives based in Chungcheong, pink plus signs), and Independents (black crosses). The configuration highlights both cross-party separation and within-party cohesion, as well as bill clusters associated with distinct issue agendas.
    }
    \label{fig:lsirm_2d}
\end{figure}

Figure~\ref{fig:lsirm_2d} displays the two-dimensional LSIRM configuration for the 17th Korean National Assembly. Legislators and bills are jointly embedded in a shared Euclidean space, with legislators colored by party affiliation and bills shown as gray points. The parties with well-documented ideological distinctiveness---the Uri Party (UP), the Grand National Party (GNP), and the Democratic Labor Party (DLP)---occupy clearly separated regions of the space, indicating that their roll-call behavior cannot be reduced to a single left--right axis.

Despite this divergence across the three major parties, each bloc exhibits strong internal cohesion, with members clustering tightly around their party-specific centroid. Smaller parties, including DP, LDU, and Independents, are positioned between UP and GNP, suggesting that they do not systematically deviate from the dominant partisan structure in ways that would generate an additional ideological dimension. The resulting configuration confirms that the legislative landscape of the 17th Korean National Assembly is fundamentally multidimensional: the primary axis separates the governing UP from the opposition GNP, while the secondary axis distinguishes the DLP from both major blocs.

The joint legislator--bill embedding also reveals that bills are not uniformly scattered throughout the space. Instead, they divide into two clusters: both lie relatively close to the UP--GNP region, but one cluster is positioned substantially farther from the DLP than the other. This pattern indicates that bill content associated with DLP-salient issues plays a substantial role in shaping the geometry of roll-call divisions. By embedding legislators and bills within a common Euclidean space, LSIRM allows ideological positions to be interpreted through the lens of neighboring bills. This geometric structure provides the foundation for our subsequent analysis, in which we use topic-based beta regression to quantify how legislators' proximity to issue-specific bill clusters translates into measurable patterns of cohesion and polarization.

\subsection{Issue-Specific Ideological Spectra}

\begin{figure}[htbp]
    \centering

    \begin{minipage}{0.47\linewidth}
        \centering
        \includegraphics[width=\linewidth]{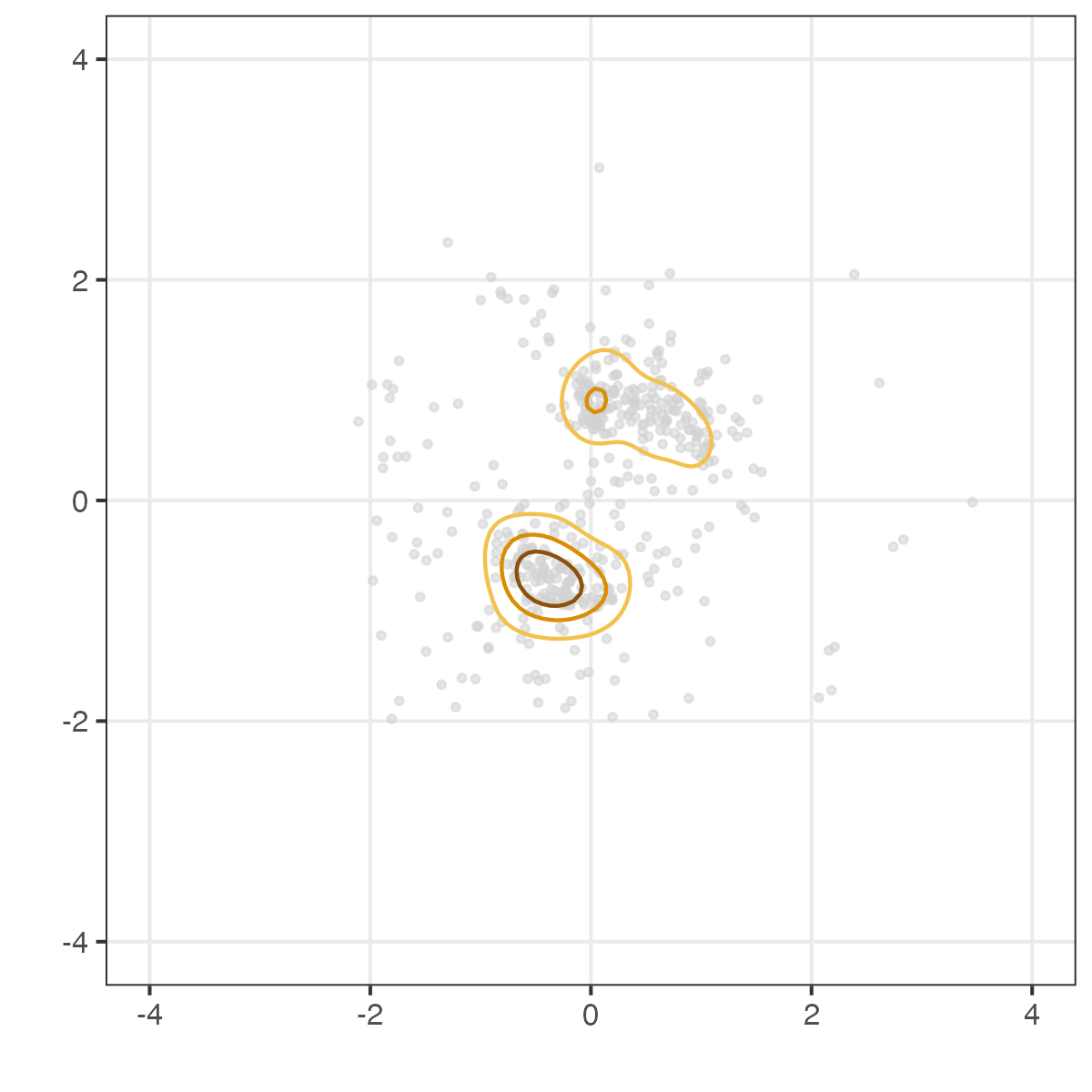}
        \small (a) Land and Regional Development
    \end{minipage}
    \hfill
    \begin{minipage}{0.47\linewidth}
        \centering
        \includegraphics[width=\linewidth]{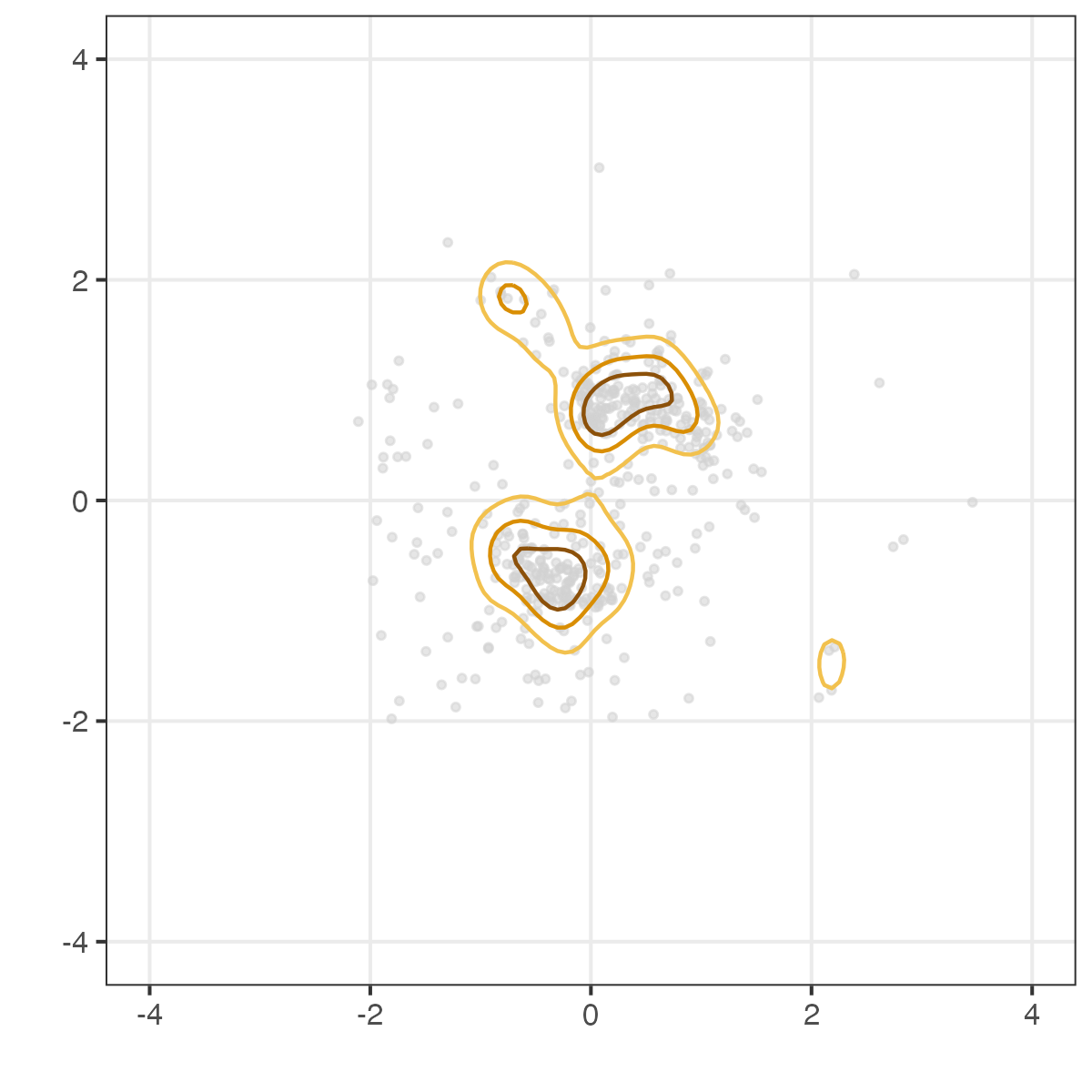}
        \small (b) Taxation
    \end{minipage}

    \vspace{2mm}

    \begin{minipage}{0.47\linewidth}
        \centering
        \includegraphics[width=\linewidth]{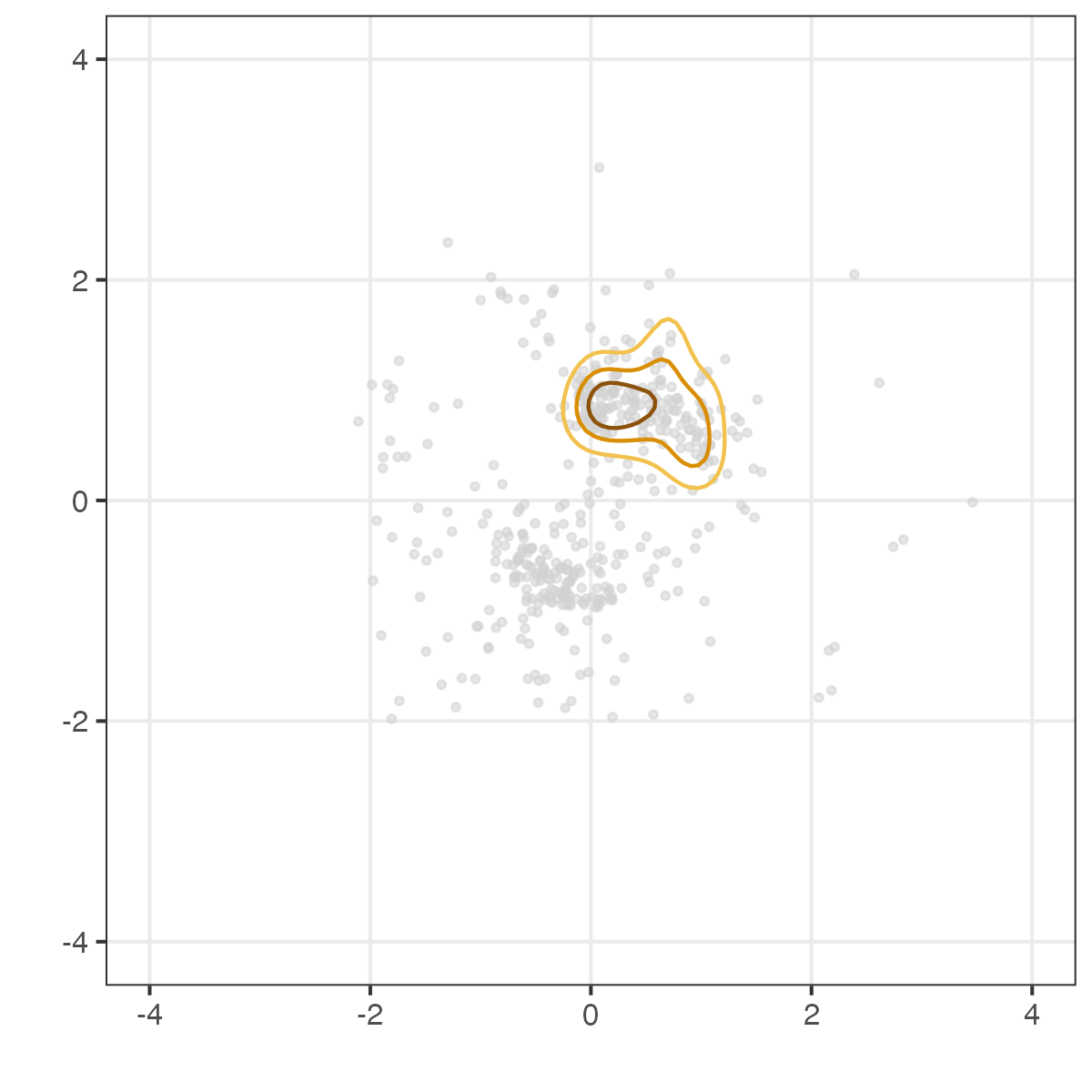}
        \small (c) Judiciary
    \end{minipage}
    \hfill
    \begin{minipage}{0.47\linewidth}
        \centering
        \includegraphics[width=\linewidth]{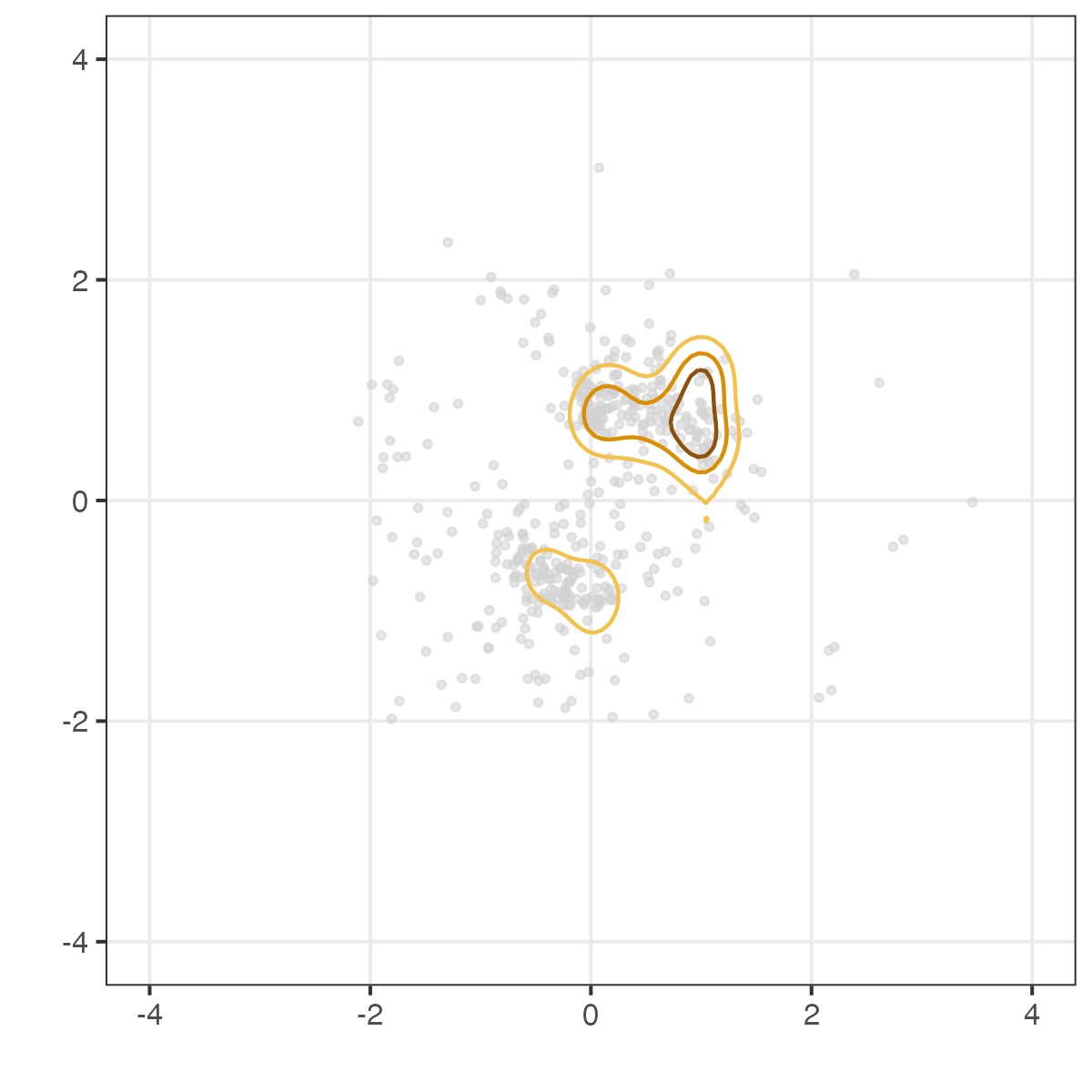}
        \small (d) Social
    \end{minipage}

    \vspace{2mm}

    \includegraphics[width=0.32\linewidth]{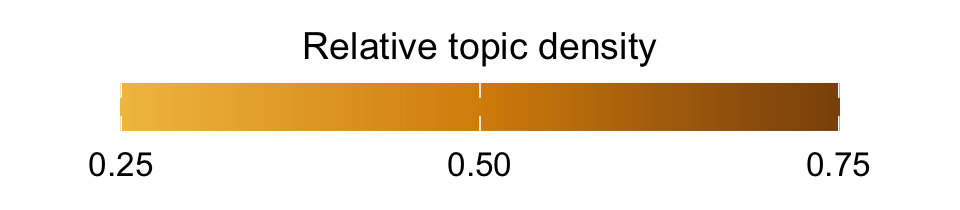}

    \caption{\textcolor{black}{Topic-weighted distributions of bills in the estimated latent space. Gray points indicate posterior mean bill positions, and contours summarize the spatial concentration of bills weighted by their estimated topic proportions. Darker contours indicate higher relative weighted density within each topic. Because densities are normalized within topic, color intensity should be interpreted within rather than directly across panels.}}
    \label{fig:bill-topic-density}
\end{figure}

\textcolor{black}{Figure~\ref{fig:bill-topic-density} illustrates how bill-level topic information is distributed over the estimated bill positions in the latent space. The weighted-density contours summarize where bills with relatively high proportions of each topic are concentrated. The four selected topics exhibit distinct spatial patterns. Land and Regional Development is concentrated primarily in the lower region of the bill space, whereas Judiciary is concentrated in the upper region. Taxation displays a more dispersed, multimodal pattern across both upper and lower regions and several peripheral areas. Social is concentrated predominantly in the upper-right region, with a much smaller secondary concentration in the lower-left. These patterns indicate that bill content is systematically associated with latent bill positions, while the overlap across topics is consistent with the mixed-membership representation of bill content. We therefore use the continuous topic proportions as bill-level covariates in the subsequent beta-regression component. Figure~\ref{fig:bill-topic-density} presents four representative topics, while the corresponding maps for the remaining four topics are reported in Section~\ref{sec:additional-bill-topic-maps} of the Supplementary Materials.}

We now turn to the issue-specific coefficients from the joint LSIRM--beta regression model. In our application, issue content is operationalized via topic proportions from BERTopic, so we report results by topic-defined issue domain. \textcolor{black}{For each legislator $i$ and issue domain $k$, the coefficient $\beta_{ik}$ measures how far the affinity of legislator $i$ toward bills emphasizing domain $k$ departs from the affinity of the same legislator toward bills emphasizing the omitted Business baseline. Comparing these coefficients across legislators and topics allows us to study between-issue variation and within-party cohesion in this Business-relative sense.}

\textcolor{black}{Table~\ref{tab:issue-variation} reports summary statistics of the posterior mean coefficients by topic, each expressed relative to the omitted Business baseline. For each issue area the first column gives the mean UP coefficient minus the mean GNP coefficient. This difference measures issue-specific inter-party polarization in affinity toward that topic relative to Business.} The second column gives the standard deviation of legislators' posterior means across the entire Assembly, capturing overall cross-legislator dispersion within that issue. The remaining columns report within-party standard deviations for UP, GNP, and DLP, respectively, which we interpret as measures of intra-party heterogeneity.

\begin{table}[htbp]
\centering
\scriptsize
\setlength{\tabcolsep}{3pt}
\renewcommand{\arraystretch}{1.25}

\begin{tabularx}{\textwidth}{
>{\raggedright\arraybackslash}p{5.8cm}
>{\centering\arraybackslash}p{2.6cm}
>{\centering\arraybackslash}p{2.0cm}
>{\centering\arraybackslash}p{1.8cm}
>{\centering\arraybackslash}p{1.8cm}
>{\centering\arraybackslash}p{1.8cm}
}
\toprule
\makecell{\textbf{Issue Area}} &
\makecell{\textbf{Party Mean}\\\textbf{Diff.\,(UP--GNP)}} &
\makecell{\textbf{\textcolor{black}{Overall}}\\\textbf{\textcolor{black}{Legislator SD}}} &
\makecell{\textbf{Within}\\\textbf{UP SD}} &
\makecell{\textbf{Within}\\\textbf{GNP SD}} &
\makecell{\textbf{Within}\\\textbf{DLP SD}} \\
\midrule
\midrule
Land and Regional Development            &  -0.075 & 0.065 & 0.050 & 0.040 & 0.004 \\
Taxation                                 &   0.362 & 0.194 & 0.084 & 0.042 & 0.011 \\
Armed Services, Patriots, and Veterans   &   0.101 & 0.188 & 0.155 & 0.077 & 0.015 \\
Judiciary                                &   0.095 & 0.226 & 0.168 & 0.097 & 0.013 \\
Local Governmental Affairs               &   0.268 & 0.194 & 0.121 & 0.046 & 0.007 \\
Grants and Local Government Budget       &   0.355 & 0.190 & 0.091 & 0.039 & 0.006 \\
Social                                   &  -0.201 & 0.202 & 0.130 & 0.094 & 0.021 \\

\bottomrule
\end{tabularx}

\caption{
\textcolor{black}{Cross-legislator and within-party variation in the posterior mean affinity coefficients, each expressed relative to the omitted Business baseline. The first column reports the mean UP coefficient minus the mean GNP coefficient and measures issue-specific inter-party polarization. The second column reports the standard deviation across all legislators and measures cross-legislator dispersion within that issue. The remaining columns report within-party standard deviations for UP, GNP and DLP and measure intra-party heterogeneity.
}}
\label{tab:issue-variation}
\end{table}

\begin{figure}[htbp]
    \centering
    \includegraphics[width=0.9\linewidth]{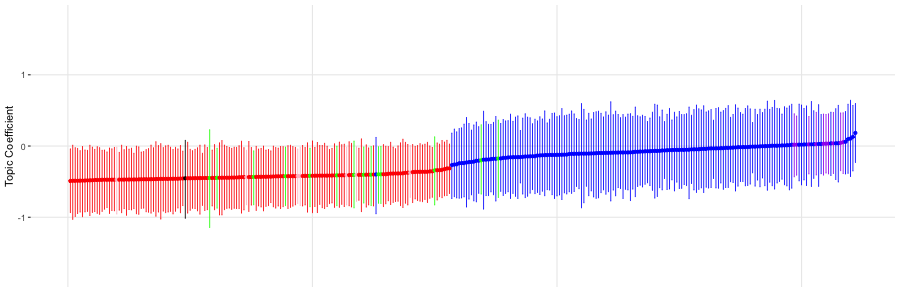}\\[-3pt]
    \text{(a) Taxation}\\[4pt]
    \includegraphics[width=0.9\linewidth]{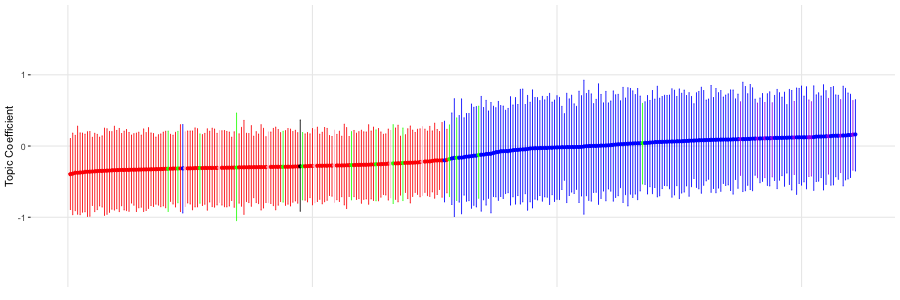}\\[-3pt]
    \text{(b) Grants and Local Government Budget}\\[4pt]
    \includegraphics[width=0.9\linewidth]{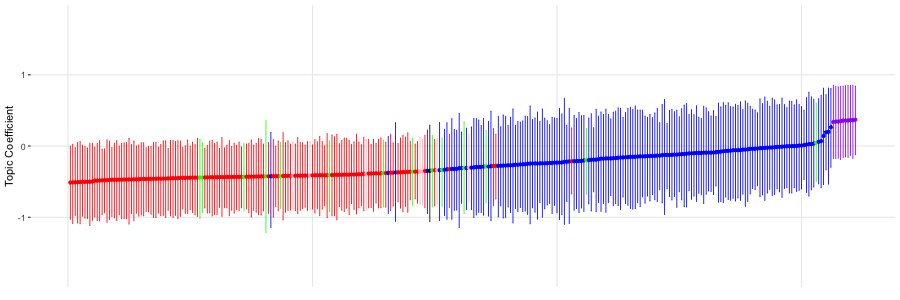}\\[-3pt]
    \text{(c) Local Governmental Affairs}\\[4pt]
    
    \caption{
    Polarized issue areas: Taxation (a), Grants and Local Government Budget (b), and Local Governmental Affairs (c). Each panel plots legislators' posterior mean coefficients with 95\% credible intervals, ordered from smallest to largest within each topic. Colors indicate party affiliation (see Figure~\ref{fig:lsirm_2d} for legend). These topics exhibit clear separation between UP and GNP; within-party dispersion varies by topic, distinguishing disciplined fiscal domains with tight clusters from more cross-cutting domains with broader within-party heterogeneity.
    }
    \label{fig:topic_polarized1}
\end{figure}

\begin{figure}[htbp]
    \centering
    \includegraphics[width=0.9\linewidth]{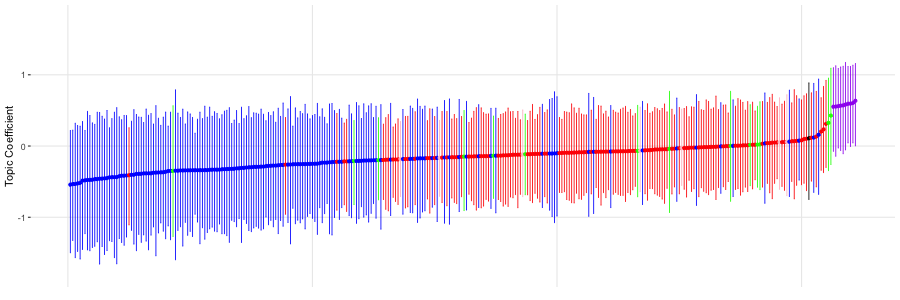}\\[-3pt]
    \text{(d) Social}\\[4pt]
    \caption{
    Polarized issue areas (continued): Social (d). Legislators' posterior mean coefficients are plotted with 95\% credible intervals, ordered from smallest to largest. Colors indicate party affiliation (see Figure~\ref{fig:lsirm_2d} for legend). Although UP and GNP are clearly separated, within-party dispersion is noticeably larger than in the fiscal domains of Figure~\ref{fig:topic_polarized1}, indicating that social issues generated meaningful heterogeneity within each major party.
    }
    \label{fig:topic_polarized2}
\end{figure}

\begin{figure}[htbp]
    \centering
    \includegraphics[width=0.9\linewidth]{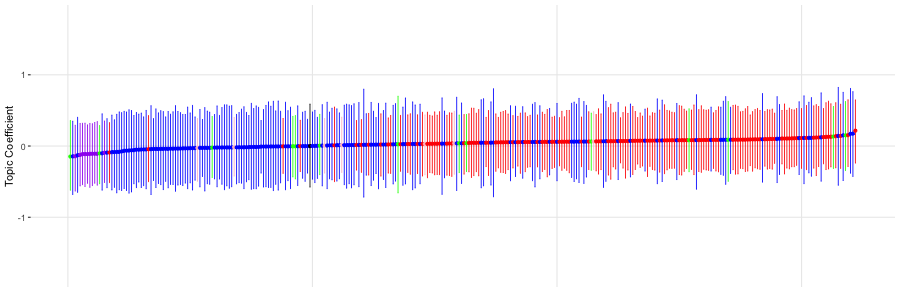}\\[-3pt]
    \text{(a) Land and Regional Development}\\[4pt]
    \includegraphics[width=0.9\linewidth]{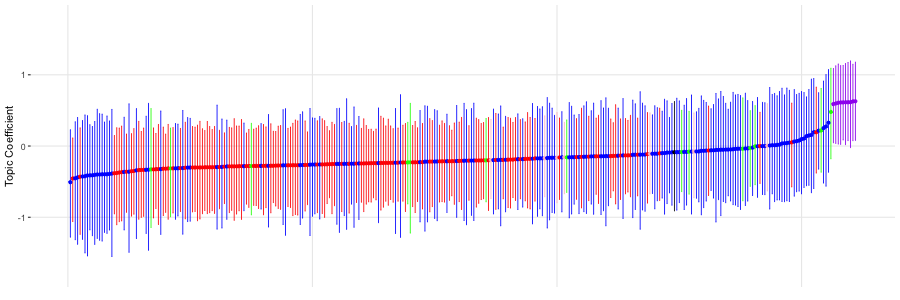}\\[-3pt]
    \text{(b) Armed Services, Patriots, and Veterans}\\[4pt]
    \includegraphics[width=0.9\linewidth]{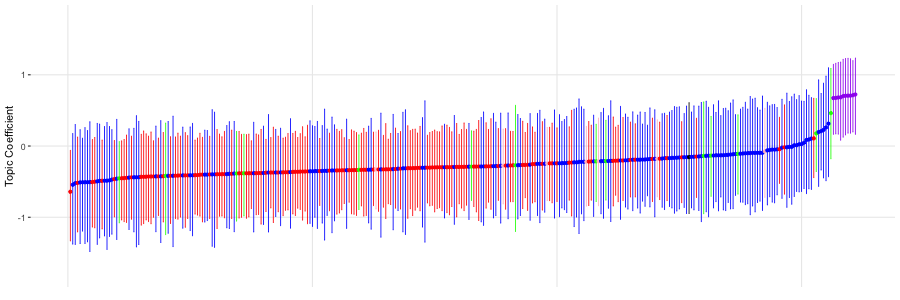}\\[-3pt]
    \text{(c) Judiciary}\\[4pt]
    
    \caption{
    Non-polarized issue areas: Land and Regional Development (a), Armed Services, Patriots, and Veterans (b), and Judiciary (c). Each panel plots legislators' posterior mean coefficients with 95\% credible intervals, ordered from smallest to largest within each topic. Colors indicate party affiliation (see Figure~\ref{fig:lsirm_2d} for legend). Land and Regional Development shows uniformly low dispersion, indicating minimal partisan conflict. By contrast, Armed Services and Judiciary exhibit substantial heterogeneity despite limited UP--GNP separation, highlighting the role of minor-party distinctiveness and within-party variation in structuring legislative conflict on these issues.
    }
    \label{fig:topic_unpolarized}
\end{figure}

To further substantiate the topic-specific patterns, Figures~\ref{fig:topic_polarized1} -- \ref{fig:topic_unpolarized} plot the distributions of legislators' posterior mean coefficients with 95\% credible intervals, ordered from smallest to largest within each issue area. We interpret these panels using three complementary criteria. First, \emph{major-party polarization} is assessed by the difference in party-level mean coefficients between UP and GNP, where larger absolute gaps indicate stronger issue-specific separation. Second, \emph{major-party cohesion} is measured by within-party dispersion (the within-UP and within-GNP standard deviations), reflecting the extent of internal discipline or factional disagreement within each major party. Third, we evaluate the \emph{role of DLP} by examining whether its mean position forms a distinct cluster and whether it lies closer to UP or GNP in party-mean distance. Visually, these criteria correspond to whether the ordered coefficients display clear discontinuities versus a smooth continuum, how steeply the ordering changes across legislators, and whether party colors form compact blocks or appear intermixed. Guided by this framework, we first separate topics into polarized versus non-polarized domains and then describe how cohesion and DLP distinctiveness refine the substantive interpretation within each domain.

\textcolor{black}{Figures~\ref{fig:topic_polarized1} and \ref{fig:topic_polarized2} present the polarized issue areas, where UP and GNP separate in their mean coefficients. Two fiscal topics, Taxation and Grants and Local Government Budget, display the strongest party-line patterns. Polarization is accompanied by limited within-party dispersion for both UP and GNP, which indicates that distributive and revenue-related legislation aligned with disciplined partisan competition. Local Governmental Affairs is also polarized, and its within-party pattern is asymmetric, with UP legislators showing broader dispersion than their GNP counterparts. The Social topic shows a different configuration. Separation between UP and GNP is present, and within-party dispersion is larger, which indicates that social issues were cross-cutting and generated heterogeneity within each major party rather than uniformly disciplined blocs.}

\textcolor{black}{Figure~\ref{fig:topic_unpolarized} shows issue areas with limited UP--GNP polarization. These topics show weaker separation along the UP--GNP cleavage, which indicates that their cross-legislator heterogeneity is not organized by the major-party divide. Land and Regional Development is the most uniform topic, with the smallest overall dispersion and low within-party variation, which indicates that this domain was not a partisan cleavage in the 17th Assembly. Armed Services, Patriots, and Veterans and Judiciary show weak separation between UP and GNP in their mean coefficients while retaining substantial dispersion across legislators. Heterogeneity in these domains is therefore not structured by the UP--GNP cleavage, which motivates closer attention to the positioning of minor parties.}

\textcolor{black}{Conventional one-dimensional scaling yields unstable and method-dependent placements of UP and DLP in this multi-party legislature, as shown in Figure~\ref{fig:17th_dim1_conventional}. It is therefore relevant to examine whether the issue-specific affinities of DLP form a coherent pattern once we move beyond a single left--right ordering. DLP has been electorally small in terms of seat share and has remained distinct from the two major parties throughout this period, which motivates examining how its issue-specific coefficients are structured.}

Across several issue areas, DLP legislators form a distinct cluster even in contexts where the two major parties are not strongly polarized. This pattern is most evident in Armed Services, Judiciary, and Local Governmental Affairs. In the first two domains, UP and GNP are relatively close in their mean positions, yet DLP's mean coefficients systematically depart from the major-party band, indicating issue-specific stances that cannot be reduced to the UP--GNP ordering. Moreover, DLP exhibits comparatively tight within-party dispersion, suggesting that these deviations reflect a coherent minor-party profile rather than idiosyncratic behavior. This finding complements the LSIRM geometry shown in Figure~\ref{fig:lsirm_2d}, where DLP occupies a clearly separated region and contributes to the separation of bill clusters. Taken together, the results imply that DLP-related voting considerations structure roll-call divisions along dimensions that are not captured by major-party polarization alone.

The Social domain further illustrates that conventional ideological labels need not map one-to-one onto issue-specific affinity dimensions. \textcolor{black}{Social legislation is itself a polarized issue area, and Table~\ref{tab:issue-variation} shows that UP and GNP mean coefficients differ substantially. Although DLP is typically characterized as left-wing, its mean position in Social legislation lies closer to GNP than to UP in party-mean distance. We report this as a descriptive feature of the estimated geometry rather than as evidence of voting alignment with either major party, since these affinities are estimated from bill-level topic covariates rather than from observed vote agreements. As an alternative construction of the bill-information matrix $\mathbf{X}$, we also estimate a committee-indicator specification, with the resulting legislator--committee affinity patterns reported in Section~\ref{supp:committee} of the Supplementary Materials.}

\subsection{Model Validation}
\subsubsection{Posterior Predictive Checks for LSIRM (Vote Outcomes)}
\begin{figure} [htb]
    \centering
    \includegraphics[width=0.46\linewidth]{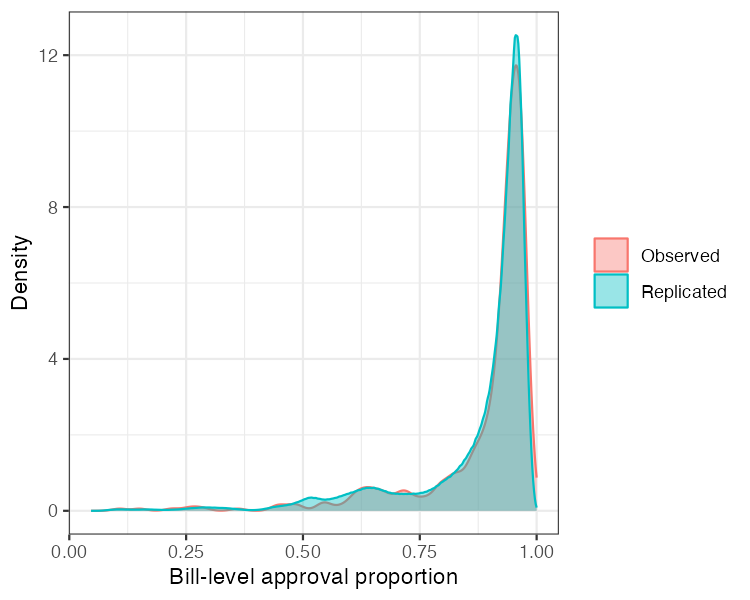}
    \includegraphics[width=0.46\linewidth]{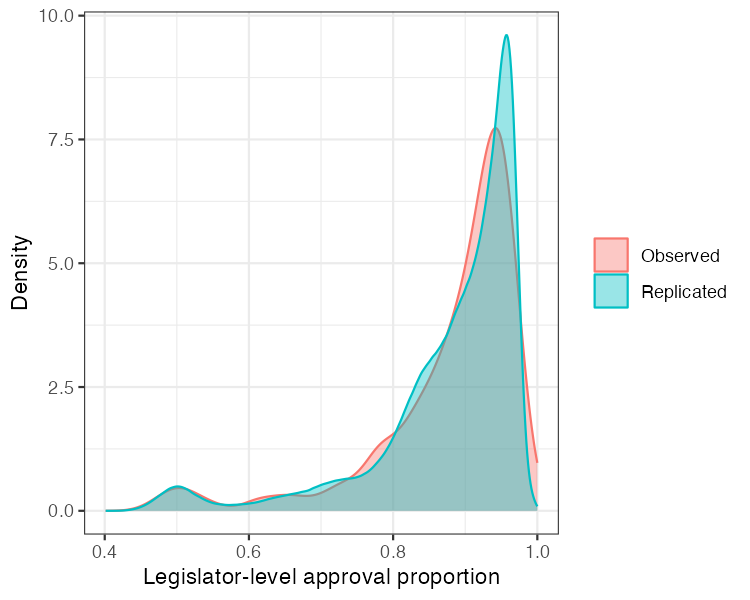}
    \caption{Posterior predictive checks for the LSIRM component of the joint model. The left panel compares observed bill-level approval proportions to replicated distributions; the right panel presents the corresponding comparison at the legislator level. Replicated distributions are generated from the LSIRM likelihood using posterior draws of latent positions and model parameters. The close alignment between observed and replicated distributions indicates good calibration of the LSIRM layer.}
    \label{fig:ppc_LSIRM}
\end{figure}

To assess the adequacy of the LSIRM component of the joint model, we conducted posterior predictive checks based on replicated roll-call matrices. Using evenly spaced posterior draws from the \textcolor{black}{Stage 1 LSIRM chain}, we simulated replicated vote matrices by drawing binary outcomes from the LSIRM likelihood conditional on the latent positions, item and legislator intercepts, and the distance scaling parameter. For each posterior draw, we generated 100 replicated vote matrices and compared their implied bill-level and legislator-level approval rates to the observed data.

\textcolor{black}{Figure~\ref{fig:ppc_LSIRM}} presents posterior predictive density plots for bill-level and legislator-level approval proportions. The replicated distributions closely match the observed ones, indicating that the LSIRM successfully captures both cross-bill and cross-legislator variation in voting behavior. In terms of interval coverage, 96.4\% of observed bill-level approval rates and 95.7\% of observed legislator-level approval rates lie within the corresponding 95\% posterior predictive intervals. These results confirm that the latent space model provides an adequate and well-calibrated representation of roll-call voting patterns in the 17th Korean National Assembly.

\subsubsection{Held-Out Vote Prediction}\label{subsec:heldout}

\textcolor{black}{To assess the empirical leverage provided by the distance-based representation, we compared the distance-only LSIRM, which excludes the subsequent beta-regression component, with a conventional two-dimensional Bayesian item response theory model \citep{Clinton2004, pscl2024}. Both models were re-estimated after withholding the same set of observed legislator--bill votes, and their predictions were evaluated on the 19,556 held-out observations. As shown in Table~\ref{tab:birt-lsirm-validation}, the distance-only LSIRM yielded higher predictive accuracy and log predictive density together with a lower Brier score than the conventional BIRT model. These results indicate that the joint representation of legislators and bills in a common Euclidean space captures voting relationships that the conventional inner-product specification does not summarize. The two models also differ in their treatment of legislator- and bill-specific baseline tendencies, and we therefore interpret this comparison as evidence for the distance-based LSIRM specification as a whole rather than attributing the predictive improvement to the distance term alone. Section~\ref{sec:birt-comparison} of the Supplementary Materials reports the corresponding two-dimensional BIRT map together with the full-data comparison and additional validation results.}

\begin{table}[htbp]
\centering
\caption{\textcolor{black}{Held-out predictive comparison of conventional BIRT and the distance-only LSIRM}}
\label{tab:birt-lsirm-validation}
\begin{tabular}{lrrrr}
\toprule
Model & Accuracy & APRE & Mean LPPD & Brier score \\
\midrule
Two-dimensional BIRT & 0.804 & -0.833 & -0.447 & 0.137 \\
Distance-only LSIRM  & 0.932 &  0.359 & -0.194 & 0.054 \\
\bottomrule
\end{tabular}
\vskip .50cm
\begin{minipage}{0.92\linewidth}
\footnotesize
\textcolor{black}{
\it Note: Both models were estimated on the same training matrix and evaluated on 19,556 held-out vote observations. Higher values indicate better performance for accuracy, APRE and mean log pointwise predictive density. Lower values indicate better performance for the Brier score. The majority-rule benchmarks are computed from the training data alone. The negative APRE for the conventional BIRT model records that its accuracy falls below the bill-specific benchmark, which reflects the absence of a legislator-specific intercept in that specification.
}
\end{minipage}
\end{table}

\subsubsection{Internal Posterior Predictive Checks for Beta Regression}\label{subsec:internal_ppc_beta}

Because the beta-regression layer models transformed legislator--bill distances rather than observed outcomes, its adequacy is evaluated through internal posterior predictive checks. Conditional on a fixed posterior draw of the latent positions $(\mathbf{z},\mathbf{w})$, we treat the LSIRM-implied affinities $t_{ij}=\exp(-d_{ij})$ as the reference affinity surface and generate \textcolor{black}{200} replicated affinities $t_{ij}^{\mathrm{rep}}$ from the beta-regression likelihood using posterior draws of legislator-specific coefficients and \textcolor{black}{legislator-specific precision parameters}.

We first summarize each legislator's overall affinity toward the roll-call agenda using the statistic $t_{i\cdot} = \frac{1}{P}\sum_{j=1}^P t_{ij}$, which captures how closely legislator $i$ tends to align with the set of bills on average across issues. \textcolor{black}{Comparing the observed values of $t_{i \cdot}$ with their posterior predictive distributions---pooling replicate draws across all $m=25$ Multiple-Imputation draws of $\Theta_{\mathrm{LS}}$ (Section~\ref{sec:method}; $R=200$ replicates per draw, 5,000 pooled in total) into a single predictive band, then checking each of the 25 draws' own observed reference values against that same pooled band and averaging the resulting coverage indicator---96.1\% (SD $=1.4$ percentage points across the 25 reference draws) of legislators' observed mean affinities fall within the corresponding 95\% posterior predictive interval. Coverage is close to nominal at every interval level examined (52.8\%, 82.3\%, and 92.1\% at the nominal 50\%, 80\%, and 90\% levels, respectively). This is a direct empirical payoff of allowing $\phi_i$ to vary by legislator (Section~\ref{sec:method}): under the single shared precision considered as an alternative in Supplementary Materials Section~\ref{sec:variable-precision}, coverage at the 95\% level falls to 89.8\% and is concentrated among legislators positioned far from the centroid of the two dominant parties (Uri Party and GNP), most severely for the Democratic Labor Party (DLP)---an ideologically peripheral bloc that is nonetheless highly internally cohesive, exhibiting the lowest within-party dispersion of topic coefficients of any party in our sample. Under legislator-specific precision, DLP's miss rate (2.8\%) is no longer elevated relative to the other parties' (2.3--4.3\%), confirming that allowing $\phi_i$ to vary resolves this specific miscalibration rather than merely relocating it.} The density of the replicated legislator-level means closely overlaps the observed distribution, indicating that the beta-regression layer reproduces the overall cross-legislator variation implied by the LSIRM distances.

At the aggregation level, we additionally examine the global mean affinity, $t_{\cdot\cdot} = \frac{1}{NP}\sum_{i=1}^N \sum_{j=1}^P t_{ij}$, which summarizes the aggregate level of legislator--bill proximity in the data. The posterior predictive distribution of $t_{\cdot\cdot}$ is concentrated around the observed value, providing further evidence that the beta-regression model is well calibrated at the aggregate level.

\begin{figure}[htb]
    \centering
    \begin{minipage}{0.47\linewidth}
        \centering
        \includegraphics[width=\linewidth]{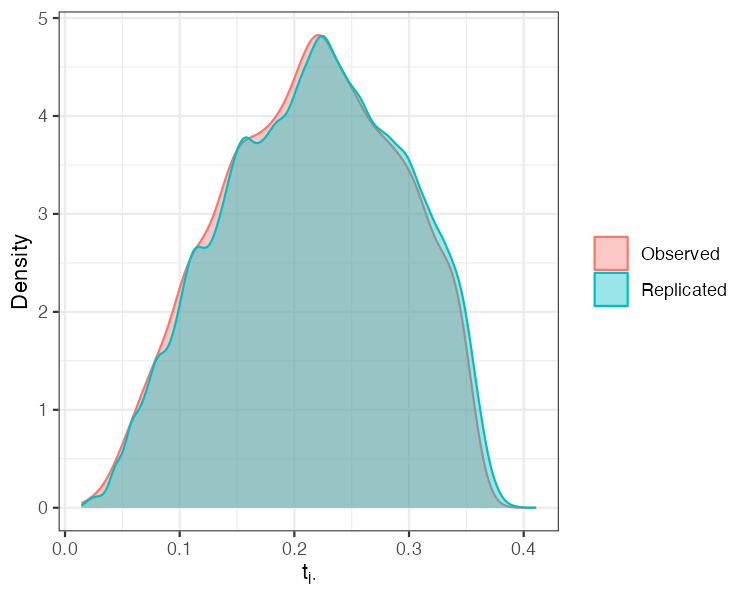}
        \text{(A)}
    \end{minipage}
    \begin{minipage}{0.47\linewidth}
        \centering
        \includegraphics[width=\linewidth]{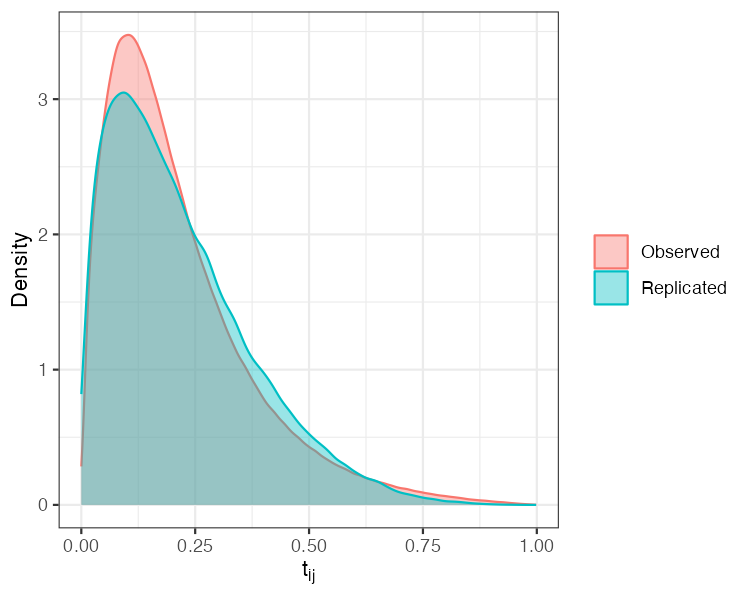}
        \text{(B)}
    \end{minipage}
    \caption{
    Posterior predictive checks for the beta-regression layer. Panel (A) compares observed and replicated legislator-level mean affinities $t_{i\cdot}$; panel (B) compares the observed and replicated global mean affinity $t_{\cdot\cdot}$. Replicated distributions are generated from the beta-regression model conditional on LSIRM-implied affinities. The close correspondence between observed and replicated quantities indicates good calibration of the beta-regression component.
    }
    \label{fig:betappc}
\end{figure}

We additionally compute legislator-specific two-sided posterior predictive p-values based on the summary statistic $T_i = t_{i\cdot}$. Let $T_i^{\mathrm{obs}}$ denote the observed value and $T_i^{\mathrm{rep}}$ its posterior predictive replicate. We define
\[
\mathrm{PPP}_i
= 2 \min\!\left\{
\Pr\!\left(T_i^{\mathrm{rep}} \ge T_i^{\mathrm{obs}}\right),
\Pr\!\left(T_i^{\mathrm{rep}} \le T_i^{\mathrm{obs}}\right)
\right\}.
\]
In practice, $\mathrm{PPP}_i$ is approximated by Monte Carlo frequencies over replicated draws of $T_i^{\mathrm{rep}}$. This quantity measures how well the model reproduces each individual legislator's average issue-weighted affinity. The resulting two-sided predictive p-values, averaged across the 25 Multiple-Imputation draws in the same way as the coverage statistics above, are broadly dispersed, with a median of 0.51 and a mean of 0.52, and only a small fraction near the boundaries of zero or one. This pattern indicates no systematic lack of fit and provides strong evidence that the beta-regression layer accurately captures legislator-level heterogeneity in the LSIRM-derived affinity structure. As a complementary check of the beta-regression specification, party-specific empirical distributions of the transformed affinities $t_{ij}$ are shown in the Supplementary Materials \textcolor{black}{(Section~\ref{sec:beta-distributional-diagnostics})}. \textcolor{black}{Supplementary Materials Section~\ref{sec:variable-precision} additionally reports the same diagnostics under a single shared precision parameter and under a party-specific precision parameter, confirming that legislator-specific precision is not merely one adequate choice among several but is specifically what closes the coverage gap documented above.}


\subsubsection{Robustness Analysis}\label{subsec:robustness_analysis}

\textcolor{black}{Beyond the choice of latent dimension (Section~\ref{sec:method}), we assessed whether the manuscript's substantive conclusions -- issue-specific major-party polarization and within-party cohesion -- depend on other analytic choices made throughout the pipeline: the treatment of recorded abstentions, the exclusion of near-unanimous roll calls from the LSIRM vote layer, the choice of omitted reference topic, the functional form used to transform latent distances into affinities, the prior specifications for the beta-regression precision and distance-effect parameters, and the topic representation used as bill-level covariates. Each check is evaluated using three complementary metrics -- legislator-level coefficient stability, the direction of the Uri Party--Grand National Party contrast, and within-party heterogeneity. Across essentially all of these checks, the substantive conclusions reported in Section~\ref{sec:results} are stable; the one partial exception is the treatment of abstentions, which has a comparatively larger effect on legislator-level coefficient magnitudes for a subset of topics, although the direction of major-party polarization is preserved. Full results for each check are reported in the Supplementary Material (Section~\ref{sec:si_robustness_metrics}).}

\section{Conclusion}\label{sec:conclusion}

This paper develops a \textcolor{black}{two-stage, cut-posterior} Bayesian framework for analyzing issue-specific legislative behavior by combining Euclidean latent-space modeling with topic-based covariates of bills. Using LSIRM to embed legislators and bills in a common geometric space and modeling the resulting distances with beta regression, the proposed approach yields issue-specific coefficients that quantify both cross-party divergence and within-party cohesion. This offers a principled way to decompose multidimensional roll-call behavior into interpretable substantive domains without discarding information.

Applied to the 17th Korean National Assembly, the method reveals substantial heterogeneity across issue areas. Fiscal domains such as Taxation and Grants and Local Government Budget exhibit sharp partisan separation with tight within-party clustering, whereas the Armed Services, Patriots, and Veterans domain shows weak party structuring and greater intra-party variability. These findings demonstrate that the model detects meaningful differences in how parties organize around specific policy areas. More broadly, the results suggest that legislative polarization is multidimensional in Korea, and that parties maintain varying degrees of coherence depending on the policy domain.

The framework also highlights several limitations that suggest directions for future work. First, the current specification does not distinguish whether the issue content is framed positively or negatively; only the magnitude of topical association (or a binary indicator) enters the regression, and incorporating sentiment or polarity-sensitive measures may refine the interpretation of topic effects. Second, incorporating richer bill metadata---such as sponsor characteristics or more sophisticated text embeddings---may further improve interpretability and predictive performance. Third, dynamic extensions tracking how legislators reconfigure across sessions would allow the model to capture the temporal evolution of issue-specific cohesion and polarization.

Despite these limitations, the proposed model provides a flexible and computationally tractable tool for linking legislative networks with substantive issue content. By bridging geometric ideal point estimation with issue-structured covariates, it contributes a novel perspective on the study of policy-specific cohesion in legislative institutions and offers a foundation for future extensions in both political methodology and statistical modeling.

\section*{Acknowledgment} 
\noindent The authors thank the editor, associate editor, and reviewers for their constructive comments. This work was partially supported by the National Research Foundation of Korea [grant number NRF-2021S1A3A2A03088949, RS-2023-00217705, RS-2024-00333701; Basic Science Research Program awarded to IHJ]. Correspondence should be addressed to Ick Hoon Jin, Department of Applied Statistics, Department of Statistics and Data Science, Yonsei University, Seoul, Republic of Korea. E-mail: ijin@yonsei.ac.kr. 

\bibliographystyle{Chicago}
\bibliography{reference}

\end{document}